\documentclass[sigconf]{acmart} 

\usepackage{graphicx}
\usepackage{balance}  

\usepackage{graphics}
\usepackage{xspace}
\usepackage[ruled, vlined, linesnumbered]{algorithm2e} 

\usepackage{booktabs}
\usepackage{soul}
\usepackage[center]{subfigure}

\usepackage{amsmath}
\usepackage{amsfonts}
\usepackage{amsbsy}
\usepackage{amsthm}
\usepackage[]{caption}
\usepackage{multirow}
\usepackage[mathscr]{eucal} 
\usepackage{mathtools}
\usepackage{verbatim}
\usepackage{tabularx}
\usepackage{enumitem}
\usepackage{array}
\usepackage{soul}

\AtBeginDocument{%
  \providecommand\BibTeX{{%
    \normalfont B\kern-0.5em{\scshape i\kern-0.25em b}\kern-0.8em\TeX}}}

\copyrightyear{2022}
\acmYear{2022}
\acmDOI{10.1145/1122445.1122456}

\setcopyright{acmcopyright}\acmConference[WWW '22]{Proceedings of the ACM Web Conference 2022}{April 25--29, 2022}{Virtual Event, Lyon, France}
\acmBooktitle{Proceedings of the ACM Web Conference 2022 (WWW '22), April 25--29, 2022, Virtual Event, Lyon, France}
\acmPrice{15.00}
\acmDOI{10.1145/3485447.3512157}
\acmISBN{978-1-4503-9096-5/22/04}

\DeclareMathOperator*{\argmax}{arg\,max}
\DeclareMathOperator*{\argmin}{arg\,min}

\begin{document}

    \title{MiDaS: Representative Sampling from Real-world Hypergraphs}
    
    \author{Minyoung Choe}
	\affiliation{%
		\institution{KAIST}
	}
	\email{minyoung.choe@kaist.ac.kr}
	
	\author{Jaemin Yoo}
	\affiliation{%
		\institution{Seoul National University}
	}
	\email{jaeminyoo@snu.ac.kr}
	
    \author{Geon Lee}
	\affiliation{%
		\institution{KAIST}
	}
	\email{geonlee0325@kaist.ac.kr}
	
	\author{Woonsung Baek}
	\affiliation{%
		\institution{KAIST}
	}
	\email{wbaek@kaist.ac.kr}
	
	\author{U Kang}
	\affiliation{%
		\institution{Seoul National University}
	}
	\email{ukang@snu.ac.kr}
	
	\author{Kijung Shin}
	\authornote{Corresponding Author}
	\affiliation{%
		\institution{KAIST}
	}
	\email{kijungs@kaist.ac.kr}
	
    \begin{abstract}
		Graphs are widely used for representing pairwise interactions in complex systems. Since such real-world graphs are large and often evergrowing, sampling a small representative subgraph is indispensable for various purposes: simulation, visualization, stream processing, representation learning, crawling, to name a few.
However, many complex systems consist of group interactions (e.g., collaborations of researchers and discussions on online Q\&A platforms), and thus they can be represented more naturally and accurately by hypergraphs (i.e., sets of sets) than by ordinary graphs.

Motivated by the prevalence of large-scale hypergraphs, we study the problem of representative sampling from real-world hypergraphs, 
aiming to answer (Q1) what a representative sub-hypergraph is and 
(Q2) how we can find a representative one rapidly without an extensive search.
Regarding Q1, 
we propose to measure the goodness of a sub-hypergraph by comparing it with the entire hypergraph in terms of ten graph-level, hyperedge-level, and node-level statistics.
Regarding Q2, we first analyze the characteristics of six intuitive approaches in $11$ real-world hypergraphs. Then, based on the analysis, we propose \methodauto, which draws hyperedges with a bias towards those with high-degree nodes.
Through extensive experiments, we demonstrate that \methodauto is
\textbf{(a) Representative}: finding overall the most representative samples among $13$ considered approaches, \textbf{(b) Fast}: several orders of magnitude faster than the strongest competitors, which performs an extensive search, and \textbf{(c) Automatic}: rapidly searching a proper degree of bias.

	\end{abstract}
	
	\newcommand\red[1]{\textcolor{red}{#1}}
\newcommand\blue[1]{\textcolor{blue}{#1}}
\newcommand\gray[1]{\textcolor{gray}{#1}}
\newcommand\kijung[1]{\textcolor{red}{[Kijung: #1]}}
\newcommand\minyoung[1]{\textcolor{blue}{#1}}
\newcommand\geon[1]{\textcolor{brown}{#1}}

\newcommand{\smallsection}[1]{{\vspace{0.02in} \noindent {{\underline{\smash{\bf #1}}}}}}
\newtheorem{obs}{\textbf{Observation}}
\newtheorem{defn}{\textbf{Definition}}
\newtheorem{thm}{\textbf{Theorem}}
\newtheorem{axm}{\textbf{Axiom}}
\newtheorem{lma}{\textbf{Lemma}}
\newtheorem{cor}{\textbf{Corollary}}
\newtheorem{problem}{\textbf{Problem}}

\newcommand\und[1]{\underline{#1}}

\newcommand{\SM}{\mathcal{M}}%
\newcommand{\SG}{\mathcal{G}}%
\newcommand{\SSS}{\mathcal{S}}%
\newcommand{\SV}{\mathcal{V}}%
\newcommand{\SE}{\mathcal{E}}%
\newcommand{\SGH}{\mathcal{\hat{G}}}%
\newcommand{\SEH}{\mathcal{\hat{E}}}%
\newcommand{\SVH}{\mathcal{\hat{V}}}%
\newcommand{\SEVH}{\SE({\SVH})}%
\newcommand{\SVEH}{\SV({\SEH})}%
\newcommand{\SD}{\mathcal{D}}%

\newcommand{\fhat}{\hat{f}}%
\newcommand{\flargehat}{\hat{F}}%
\newcommand{\yhat}{\hat{y}}%

\newcommand{\Ghat}{\SGH=(\SVH,\SEH)}%
\newcommand{\Glong}{\SG=(\SV,\SE)}%

\newcommand{\GALG}{\SGH_{ALG}}%
\newcommand{\GRNS}{\SGH_{RNS}}%
\newcommand{\GRDN}{\SGH_{RDN}}%
\newcommand{\GRHS}{\SGH_{RHS}}%
\newcommand{\GTIHS}{\SGH_{TIHS}}%
\newcommand{\cmark}{\ding{51}}%
\newcommand{\xmark}{\ding{55}}%

\newcommand{\method}{\textsc{MiDaS-Basic}\xspace}
\newcommand{\methodgrid}{\textsc{MiDaS-Grid}\xspace}
\newcommand{\methodauto}{\textsc{MiDaS}\xspace}

\definecolor{myred}{RGB}{195, 79, 82}
\definecolor{mygreen}{RGB}{86, 167 104}
\definecolor{myblue}{RGB}{74, 113 175}

\newcommand{\bigcell}[2]{\begin{tabular}{@{}#1@{}}#2\end{tabular}}

\let\oldnl\nl
\newcommand{\nonl}{\renewcommand{\nl}{\let\nl\oldnl}}
	\maketitle
    
    \section{Introduction}
	\label{sec:intro}




Complex systems consist of entities and their interactions; and graphs have been used widely to model such complex systems.
Examples include online social networks, email communication networks, hyperlink networks, and internet router networks. 
As a result of the emergence and development of the World Wide Web and its applications, large-scale graphs have become prevalent.


Since it is time-consuming and often practically impossible to collect and analyze every component in such large-scale graphs,
sampling has been employed for various tasks on graphs, including:

\noindent $\circ$ \textbf{Simulation}: Packet-level simulations of internet protocols are time-consuming, and they need to be repeated multiple times to confirm the reliability of protocols. In order to reduce computation time, samples of internet topologies have been utilized~\cite{krishnamurthy2007sampling,krishnamurthy2005reducing}.

\noindent $\circ$ \textbf{Visualization}: Visualizing a large graph is challenging due to the vast number of components (i.e., nodes and edges), the lack of screen space, and the complexity of layout algorithms. 
    A small representative subgraph can be used to mitigate these difficulties. The subgraph can be set to include important nodes~\cite{kurant2012coarse, gilbert2004compressing} or to maximize the visualization quality measures \cite{cui2006measuring, bertini2011improving}.

\noindent $\circ$ \textbf{Stream Processing}: A dynamic graph that grows indefinitely is naturally treated as a stream of edges whose number can potentially be infinite. Thus, only a subgraph, instead of the entire graph, is maintained for detecting
    outliers~\cite{aggarwal2011outlier, eswaran2018sedanspot}, predicting edges~\cite{zhao2016link}, and counting triangles~\cite{tsourakakis2009doulion, stefani2017triest, lee2020temporal, lim2015mascot, ahmed2014graph} in such a stream. 

\noindent $\circ$  \textbf{Crawling}: Online social networks (e.g., Facebook and Twitter)  provide information on connections mainly by API queries. Limitations on API request rates make it inevitable to deal with a subgraph instead of the entire graph~\cite{xu2014general,katzir2011estimating, gjoka2010walking, gjoka2011practical, lee2012beyond}.
      
\noindent $\circ$   \textbf{Graph Representation Learning}: Despite their wide usage, graph neural networks (GNNs) often suffer from scalability issues due to the recursive expansion of neighborhoods across layers. Sampling has been employed to accelerate training by limiting the size of the neighborhoods. To this end,  nodes~\cite{chen2018fastgcn}, neighborhoods~\cite{hamilton2017inductive, chen2017stochastic, zou2019layer}, or subgraphs~\cite{zeng2019graphsaint, chiang2019cluster} can be sampled.

\begin{figure}[t]
    \begin{minipage}{0.57\linewidth}
        \includegraphics[width=1.0\textwidth]{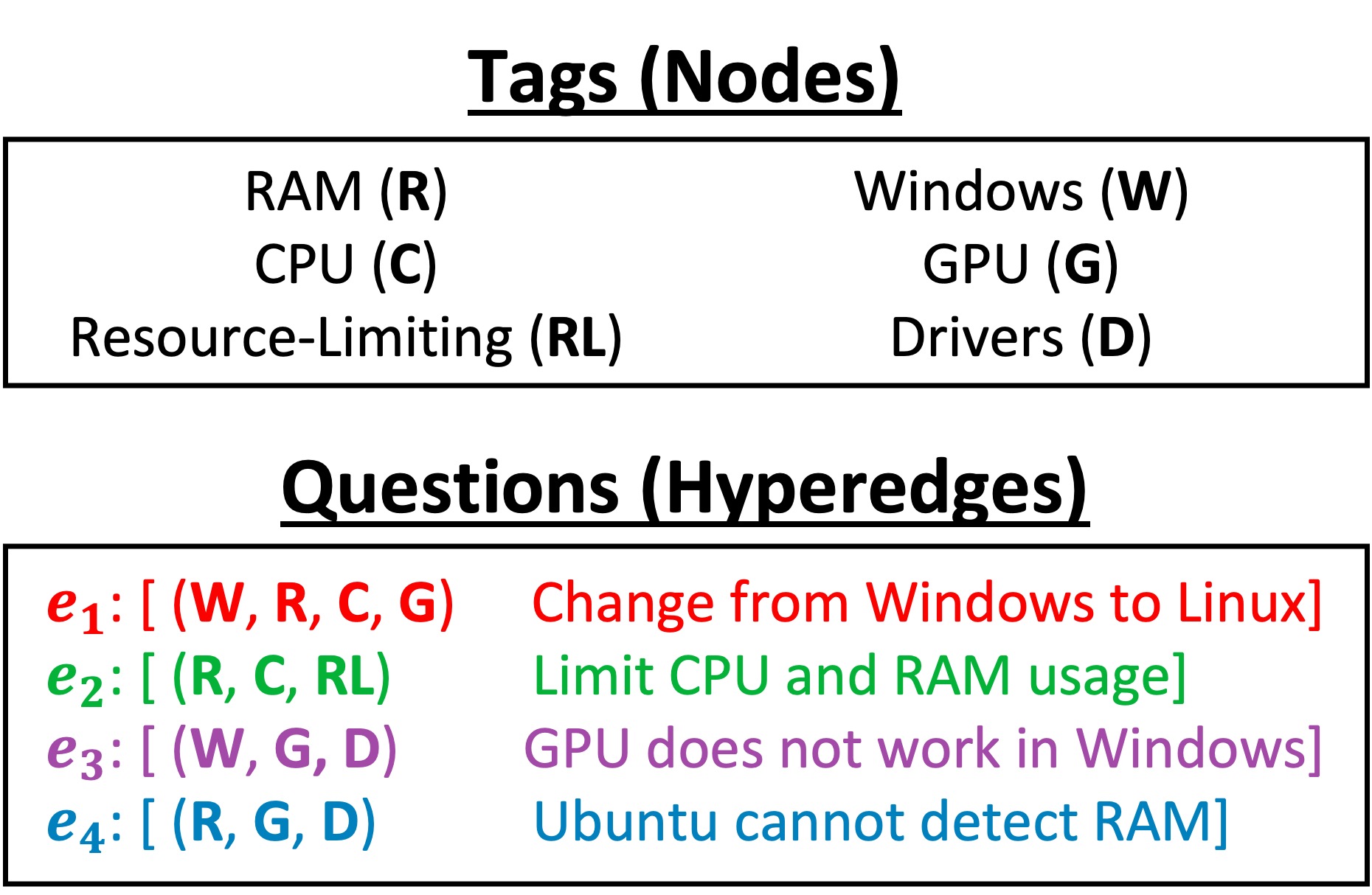}
    \end{minipage}
    \begin{minipage}{0.31\linewidth}
        \includegraphics[width=1.0\textwidth]{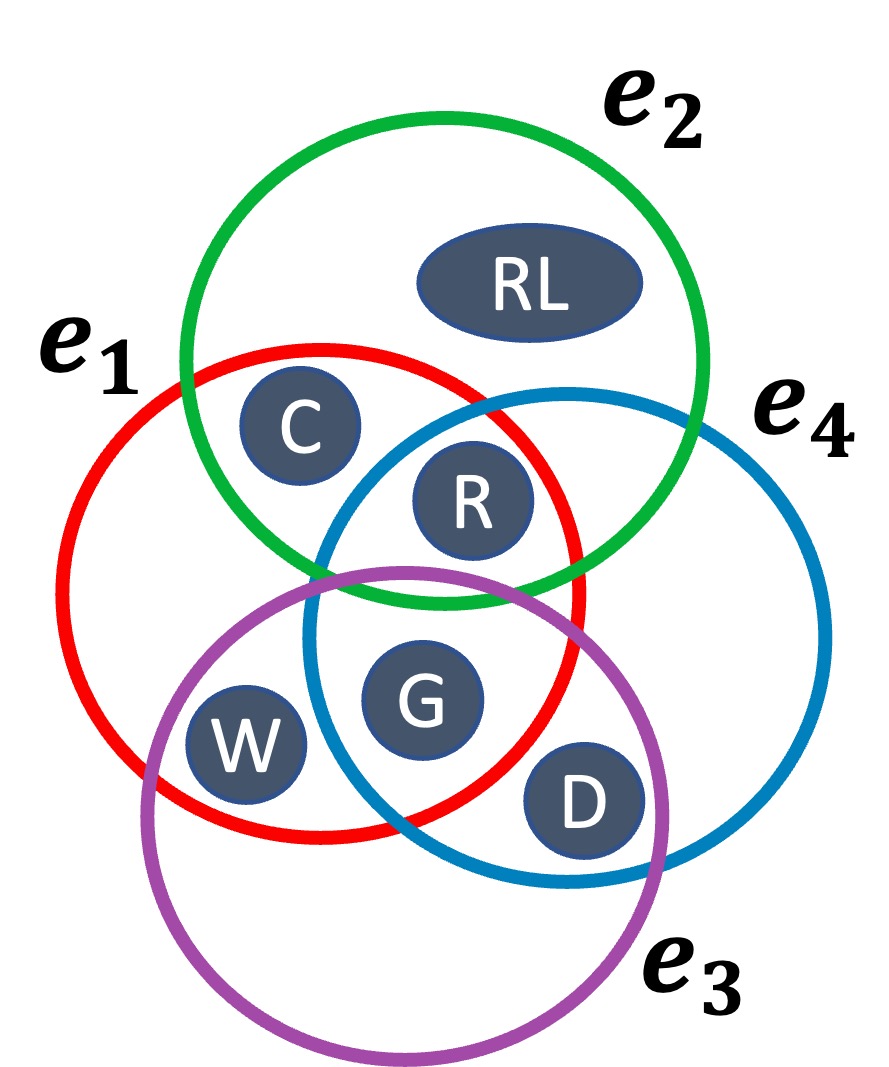}
    \end{minipage}
    
    \begin{minipage}{1.0\linewidth} \small
        \hspace{0.05\linewidth} (a) Example data: tags attached to questions \hspace{0.03\linewidth} (b) Hypergraph
    \end{minipage}
    
    \caption{\label{fig:intro_ex} The questions with tags in (a) are represented as a hypergraph with six nodes and four hyperedges in (b).}
\end{figure}

Beyond pairwise interactions, group interactions exist in many complex systems, including collaborations of researchers, group discussion on online Q\&A platforms, and interactions of ingredients in recipes.
Such complex systems are more accurately represented as hypergraphs than as graphs.
A \textit{hypergraph} consists of nodes and hyperedges, and each \textit{hyperedge} is an any-size subset of nodes (see Figure~\ref{fig:intro_ex} for an example).
Modeling complex systems as hypergraphs, rather than  graphs, can help capture domain-specific structural patterns~\cite{lee2020hypergraph}, predict interactions~\cite{yoon2020much}, cluster nodes~\cite{wolf2016advantages}, and measure node importance~\cite{chitra2019random}.


In this paper, we introduce the \textit{representative hypergraph sampling problem}, where the objective is to sample a representative sub-hypergraph that accurately preserves the characteristics of a given hypergraph.
Since real-world hypergraphs are similar in size to and more complex than real-world graphs, sampling from hypergraphs provides substantial benefits, including those listed above.
Regarding the problem, we aim to answer the following questions:
\begin{itemize}[leftmargin=*]
    \item \textbf{Q1.} What is a `representative' sample? How can we measure the quality of a sub-hypergraph as a representative sample?
    \item \textbf{Q2.}  What are the benefits and limitations of simple and intuitive approaches for representative hypergraph sampling?
    \item \textbf{Q3.}  How can we  find a representative sub-hypergraph rapidly without  extensively exploring the search space?
\end{itemize}

Regarding Q1, we measure the difference between sampled and entire hypergraphs using ten statistics. Specifically, as node-level and hyperedge-level statistics, we compare the distributions of node degrees, hyperedge sizes, intersection sizes \cite{kook2020evolution}, and node-pair degrees \cite{lee2021hyperedges} in sampled and entire hypergraphs.
We also compare their average clustering coefficient \cite{do2020structural}, density \cite{hu2017maintaining}, overlapness \cite{lee2021hyperedges}, and effective diameter \cite{leskovec2005graphs, kook2020evolution} as graph-level statistics.


\begin{figure}[t]
    \vspace{-4mm}
    \centering
    
    \begin{minipage}{0.305\linewidth}
        \includegraphics[width=1.0\textwidth]{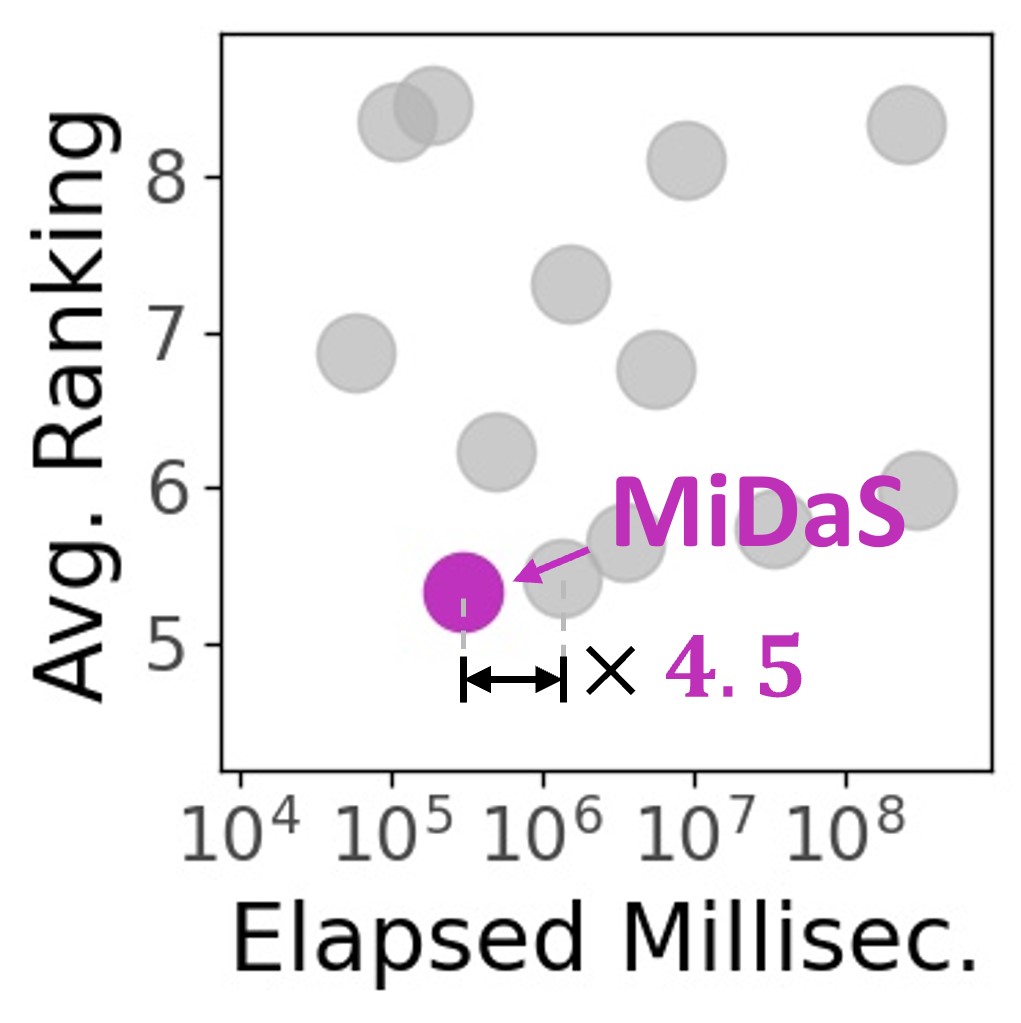} 
    \end{minipage}
    \hspace{1mm}
    \begin{minipage}{0.365\linewidth}
        \includegraphics[width=1.0\textwidth]{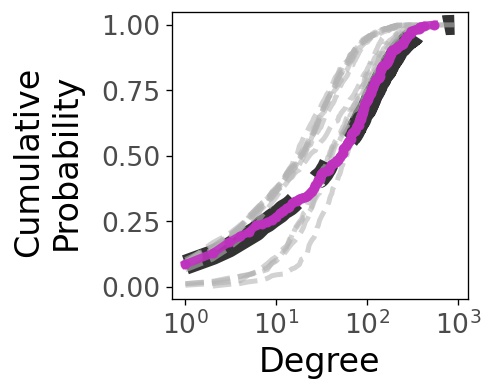}
    \end{minipage}
    \hspace{1.5mm}
    \begin{minipage}{0.22\linewidth}
        \includegraphics[width=1.0\textwidth]{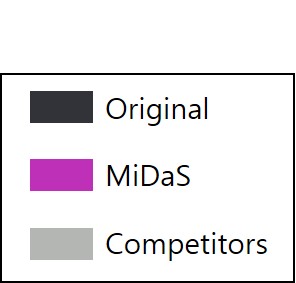}
        \vspace{3mm}
    \end{minipage}
    
    \begin{minipage}{1.0\linewidth} \small
       \hspace{0.02\linewidth} (a) Accuracy and Speed  \hspace{0.05\linewidth} (b) Degree Distribution
    \end{minipage}
    
    \caption{\label{fig:strength}Strengths of \methodauto. (a) It rapidly finds overall the most representative sub-hypergraphs among $\bf{13}$ approaches from $\bf{11}$ real-world hypergraphs. (b) Especially, it accurately preserves the degree distribution.
    See Section~\ref{sec:evaluation} for details.}
    
\end{figure}

Concerning Q2, we try six simple and intuitive sampling approaches for representative sampling from $11$ real-world hypergraphs. Then, we analyze their benefits and limitations. While some approaches preserve certain structural properties well, none of them succeeds in preserving all ten properties, demonstrating the difficulty of the considered problem.


With respect to Q3, we propose \methodauto (\underline{\smash{Mi}}nimum \underline{\smash{D}}egree Bi\underline{\smash{a}}sed \underline{\smash{S}}ampling of Hyperedges) for representative sub-hypergraph sampling. 
\method is inspired by two facts: (a) all the simple approaches fail to preserve degree distributions well and (b) the ability to preserve degree distributions is strongly correlated to the abilities to preserve other properties.
In a nutshell, \methodauto draws hyperedges with a bias towards those with high-degree nodes, and it automatically controls the degree of bias.
Through extensive experiments, we show that \methodauto performs best overall among $12$ competitors in $11$ real-world hypergraphs, as shown in Figure~\ref{fig:strength}. 

Our contributions are summarized as follows:
\begin{itemize}[leftmargin=*]
    \item \textbf{New Problem}: We deal with representative sampling from real-world hypergraphs for the first time to the best of our knowledge.
    \item \textbf{Findings}: We analyze six intuitive approaches on $11$ datasets with a focus on their limitations in preserving $10$ properties. 
    \item \textbf{Algorithm}: We propose \methodauto, which rapidly finds overall the most representative sample among $13$ methods (see Figure~\ref{fig:strength}).
\end{itemize}
\textbf{The code and datasets} are available at \cite{online2021appendix} for reproducibility.

In Section \ref{sec:problem_formulation}, we formulate our problem. In Section \ref{sec:baseline}, we analyze six intuitive sampling methods in $11$ real-world hypergraphs. Based on them, in Section \ref{sec:proposed_approach}, we propose our sampling method, \methodauto. In Section \ref{sec:evaluation}, we present experimental results. In Section \ref{sec:related}, we discuss related work. In Section \ref{sec:conclusion}, we offer conclusions.

	
	\section{Problem Formulation}
	\label{sec:problem_formulation}
	\begin{table}[t!]
    \vspace{-3mm}
	\caption{\label{tab:notations}Frequently-used symbols.}
	\scalebox{0.77}{
		\begin{tabular}{c|l}
			\toprule
			\textbf{Notation} & \textbf{Definition}\\
			\midrule
			$\Glong$ & hypergraph with nodes $\SV$ and hyperedges $\SE$\\
			$\Ghat$ & a subhypergraph of $\SG$ (i.e., $\SVH \subseteq \SV$ and $\SEH \subseteq \SE$)\\
			$p$ & target sampling portion (i.e., the ratio of sampled hyperedges) \\	
			$d_{v}$ & degree of a node $v$ in $\SG$ (i.e., the number of hyperedges including $v$) \\
			\midrule
			$\omega(e)$ & weight of hyperedge $e$ \\
			$\alpha$ & degree of bias (i.e., exponent of hyperedge weights in \methodauto)\\
			\bottomrule
		\end{tabular}}
\end{table}

In this section, we formally define the representative hypergraph sampling problem. To this end, we first introduce basic concepts and ten statistics for comparing hypergraphs of different sizes.

%


\subsection{Basic Concepts and Notations}
\label{problem_form:concepts}
See Table~\ref{tab:notations} for frequently-used symbols.
A \textit{hypergraph} $\Glong$ consists of a set of nodes $\SV$ and a set of hyperedges $\SE \subseteq2^{\SV}$. Each hyperedge $e\in \SE$ is a non-empty subset of $\SV$.
The \textit{degree} of a node $v$ is the number of hyperedges containing $v$, i.e., $d_{v}:=|\{ e \in \SE: v \in e \}|$.
A \textit{sub-hypergraph} $\Ghat$ of $\Glong$ is a hypergraph consisting of nodes $\SVH\subseteq \SV$ and hyperedges $\SEH\subseteq\SE$. 

\subsection{Statistics for Structure of Hypergraphs} \label{problem_form:properties}

We introduce ten node-level (\textbf{P1}, \textbf{P3}), hyperedge-level (\textbf{P2}, \textbf{P4}), and graph-level (\textbf{P5}-\textbf{P10}) statistics that have been used extensively for structure analysis of real-world graphs \cite{leskovec2006sampling,leskovec2005graphs} and hypergraphs \cite{do2020structural,kook2020evolution,lee2021hyperedges}.
They are used throughout this paper to measure the structural similarity of hypergraphs.


\noindent $\circ$ \textbf{P1. Degree:} We consider the degree distribution of nodes. The distribution tends to be heavy-tailed in real-world hypergraphs but not in uniform random hypergraphs~\cite{do2020structural, kook2020evolution}.

\noindent $\circ$ \textbf{P2. Size:} We consider the size distribution of hyperedges, which is shown to be heavy-tailed in real-world hypergraphs~\cite{kook2020evolution}.

\noindent $\circ$ \textbf{P3. Pair Degree:} 
We consider the pair degree distribution of neighboring node pairs. The \textit{pair degree} of two nodes is defined as the number of hyperedges containing both. 
The distribution reveals structural similarity between nodes, and it tends to have a heavier tail in real-world hypergraphs than in randomized ones~\cite{lee2021hyperedges}.

\noindent $\circ$ \textbf{P4. Intersection Size (Int. Size):} We consider the intersection-size (i.e., count of common nodes) distribution of overlapping hyperedge pairs. 
The distribution from pairwise connections between hyperedges is heavy-tailed in many real-world hypergraphs~\cite{kook2020evolution}.

\noindent $\circ$ \textbf{P5. Singular Values (SV):} We consider the relative variance explained by singular vectors of the incidence matrix. In detail, for each $i\in\{1,\cdots,R\}$, we compute $s_{i}^{2}$ / $\sum_{k=1}^{R} s_{k}^{2}$ where $s_i$ is the $i$-th largest singular value and $R$ is the rank of the incidence matrix.
Singular values indicate the variance explained by the corresponding singular vectors \cite{wall2003singular}, and they are highly skewed in many real-world hypergraphs \cite{kook2020evolution}. They are also equal to the square root of eigenvalues of weighted adjacency matrix.
For the large datasets from the threads and co-authorship domains, we use $300$ instead of $R$, and for a sample from them, we use $300/R$ of the rank of its incidence matrix.
%


\noindent $\circ$ \textbf{P6. Connected Component Size (CC):} We consider the portion of nodes in each $i$-th largest connected component in the clique expansion.
The \textit{clique expansion} of a hypergraph $\Glong$ is the undirected graph obtained by replacing each hyperedge $e\in \SE$ with the clique with the nodes in $e$.
In many real-world hypergraphs, a majority of nodes belong to a few connected components \cite{do2020structural}.

\noindent $\circ$ \textbf{P7. Global Clustering Coefficient (GCC):} We estimate the average of the clustering coefficients of all nodes in the clique expansion (defined in P6) using \cite{seshadhri2014wedge}. This statistic measures the cohesiveness of connections, and it tends to be larger in real-world hypergraphs than in uniform random hypergraphs \cite{do2020structural}.

\noindent $\circ$ \textbf{P8. Density:} The \textit{density} is defined as the ratio of the hyperedge count over the node count (i.e., $|\SE|/|\SV|$) \cite{hu2017maintaining}. Hypergraphs from the same domain tend to share a similar significance of density~\cite{lee2021hyperedges}.

\noindent $\circ$ \textbf{P9. Overlapness:} The \textit{overlapness} of a hypergraph $\Glong$ is defined as $\sum_{e \in \SE} |e| / |\SV|$. It measures the degree of hyperedge overlaps, satisfying desirable axioms \cite{lee2021hyperedges}. Hypergraphs from the same domain tend to share a similar significance of overlapness~\cite{lee2021hyperedges}.


\noindent $\circ$ \textbf{P10. Diameter:} The \textit{effective diameter} is defined as the smallest $d$ such that the paths of length at most $d$ in the clique expansion (defined in P6) connect 90\% of reachable pairs of nodes  \cite{leskovec2005graphs}. It measures how closely nodes are connected.  The effective diameter tends to be small in real-world hypergraphs \cite{kook2020evolution}.

\subsection{Problem Definition}
\label{problem_form:problem}

Based on the statistics, we formulate the representative hypergraph sampling problem in Problem~\ref{problem}.

\vspace{0.5mm}
\noindent\fbox{%
        \parbox{0.98\columnwidth}{%
        \vspace{-2mm}
\begin{problem}[Representative Hypergraph Sampling\label{problem}] \
\begin{itemize}[leftmargin=*]
    \item \textbf{Given}: - a large hypergraph $G = (\SV, \SE)$
    \item[] \qquad \quad - a sampling portion $p\in(0,1)$
    \item \textbf{Find}: a sub-hypergraph $\hat{G}=(\SVH, \SEH)$ where $\SVH\subseteq \SV$ and $\SEH\subseteq \SE$
    \item \textbf{to Preserve}: ten structural properties of $G$ measured by \textbf{P1}-\textbf{P10}
    \item \textbf{Subject to}: $|\SEH| = \lfloor |\SE| \cdot p \rfloor$
\end{itemize}
\end{problem}
        \vspace{-2mm}
        }%
    }
\vspace{0.5mm}
We evaluate the goodness of a sub-hypergraph $\SGH$ as a representative sample of $\SG$ by measuring
how precisely $\SGH$ preserves the structural properties of  $\SG$ in ten aspects.
Specifically, for each of \textbf{P1}-\textbf{P6}, which are (probability density) functions, we measure the Kolmogorov-Smirnov \textit{D-statistic}. Specifically, for functions $f$ from $\SG$ and $\fhat$ from $\SGH$, we consider their cumulative sums $F$ and $\flargehat$,\footnote{That is, $F(x):=\sum_{i=1}^{x}f(i)$ and $\flargehat(x):=\sum_{i=1}^{x}\fhat(i)$.} and we measure $\max_{x\in \SD} \{| \flargehat(x) - F(x) | \}\in[0,1]$, where $\SD$ is the domain of $f$ and $\fhat$.
For each of \textbf{P7}-\textbf{P10}, which are scalars, we measure the relative difference. Specifically, for scalars $y$ from $\SG$ and $\yhat$ from $\SGH$, we measure $|y-\yhat|/|y|$.

In Problem~\ref{problem}, the objective is to find the most representative sub-hypergraph composed of a given portion of hyperedges. However, solving to optimality is challenging as the ten structural properties need to be considered simultaneously.
In this paper, we focus on developing heuristics that work well in practice.
	
	\section{Simple and Intuitive Approaches}
	\label{sec:baseline}
\begin{table*}[t!]
\vspace{-3mm}
\centering
\caption{\label{tab:baseline} Six intuitive sampling methods are compared as described in Section~\ref{sec:observations:evaluation}.  (a) \textit{RHS} provides overall the most representative sub-hypergraphs. (b)  However, not every property is accurately preserved by \textit{RHS}.
See Section~\ref{sec:observations} for detailed analysis.
}

    {\small (a) Rankings and Z-Scores (in parentheses) averaged over five sampling portions ($10\%,\cdots,50\%$) and the $11$ datasets. Best results are in \textbf{bold}.}

    \scalebox{0.79}{
        \begin{tabular}{c cccccccccc | c}
            \toprule
            \textbf{Algorithm} & \textbf{Degree} & \textbf{Int. Size} & \textbf{Pair Degree} & \textbf{Size} & \textbf{SV} & \textbf{CC} & \textbf{GCC} & \textbf{Density} & \textbf{Overlapness} & \textbf{Diameter} & \textbf{AVG} \\
            
            
            
            \hline
            \textit{RNS} & 3.51 (-0.11) & 4.42 (0.69) & 3.95 (0.26) & 5.85 (1.54) & 2.89 (-0.24) & 3.71 (0.61) & 3.31 (0.06) & 3.07 (\textbf{-0.38}) & 3.98 (0.16) & 4.64 (0.72) & 3.93 (0.33) \\
            \textit{RDN} & \textbf{3.16} (-0.14) & 3.49 (-0.06) & 4.11 (0.30) & 4.20 (0.08) & 4.07 (0.30) & 2.25 (-0.17) & 4.60 (0.53) & 4.20 (0.35) & 3.49 (-0.09) & 3.02 (-0.21) & 3.66 (0.09) \\
            \textit{RW} & 3.96 (0.29) & 4.00 (0.15) & \textbf{2.69} (-0.28) & 4.11 (0.20) & 3.64 (0.21) & 3.73 (0.46) & 3.15 (-0.25) & 3.24 (-0.08) & 3.80 (0.29) & 3.42 (-0.15) & 3.57 (0.08) \\
            \textit{FF} & 3.89 (0.24) & 4.29 (0.41) & 3.93 (0.13) & 2.45 (-0.57) & 4.56 (0.59) & 3.09 (-0.04) & 4.16 (0.16) & 3.42 (-0.00) & 3.24 (-0.20) & 3.20 (-0.26) & 3.62 (0.05) \\
            \textit{RHS} & 3.24 (-0.13) & \textbf{1.07} (\textbf{-1.14}) & 3.64 (0.01) & \textbf{1.00} (\textbf{-1.11}) & \textbf{1.78} (\textbf{-0.96}) & \textbf{1.87} (\textbf{-0.71}) & \textbf{2.84} (\textbf{-0.30}) & 4.18 (0.44) & 3.71 (0.20) & 3.98 (0.29) & \textbf{2.73} (\textbf{-0.34}) \\
            \textit{TIHS} & 3.24 (\textbf{-0.15}) & 3.73 (-0.04) & \textbf{2.69} (\textbf{-0.43}) & 3.38 (-0.14) & 3.45 (0.10) & 2.65 (-0.15) & 2.95 (-0.21) & \textbf{2.89} (-0.33) & \textbf{2.78} (\textbf{-0.38}) & \textbf{2.75} (\textbf{-0.39}) & 3.05 (-0.21) \\
            \bottomrule
        \end{tabular}}

    \vspace{1mm}
 {\small (b) Representative results when the sampling portion is $0.3$. We provide the results regarding 6 out of 10 statistics here, and the full results can be found in \cite{online2021appendix}.}
  \vspace{1mm}
    \includegraphics[width=0.75\textwidth]{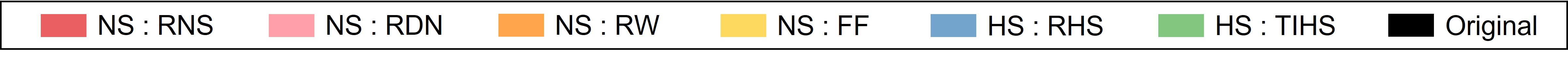} \\
    \scalebox{0.82}{
    \begin{tabular}{c|cccccc}
     \toprule
     {Data} & \textbf{\ \ \ \ \ \ \ \ \ \ Degree} & \textbf{\ \ \ \ \ \ \ \ \ \ Size} & \textbf{\ \ \ \ \ \ \ \ \ \ CC} & \textbf{\ \ \ \ \ \ \ \ \ \ Density} & \textbf{\ \ \ \ \ \ \ \ \ \ Diameter} & \textbf{\ \ \ \ \ \ \ \ \ \ GCC} \\ 
    \hline
    
    {\multirow{6}{*}{email-Eu}} & 
    \raisebox{-.9\totalheight}{\includegraphics[width=0.16\textwidth]{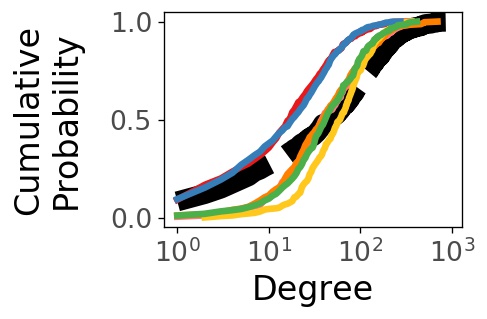}} & 
    \raisebox{-.9\totalheight}{\includegraphics[width=0.16\textwidth]{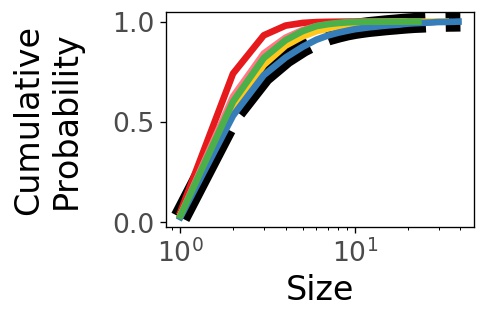}} &
    \raisebox{-.9\totalheight}{\includegraphics[width=0.16\textwidth]{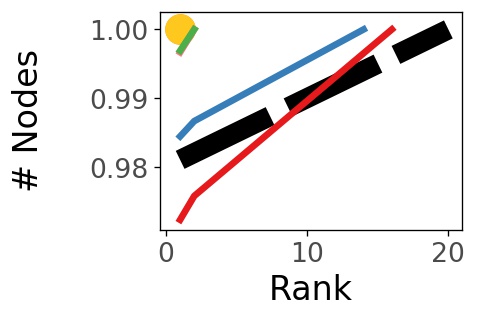}} &
    \raisebox{-.9\totalheight}{\includegraphics[width=0.16\textwidth]{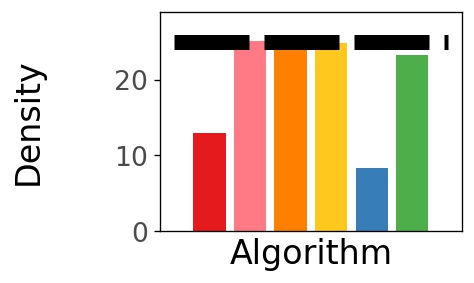}} &
    \raisebox{-.9\totalheight}{\includegraphics[width=0.16\textwidth]{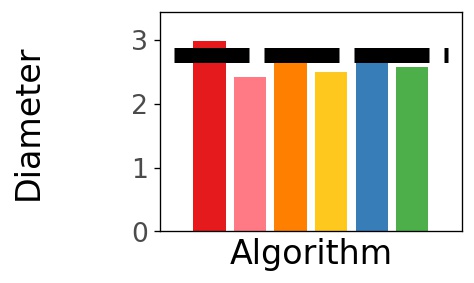}} & 
    \raisebox{-.9\totalheight}{\includegraphics[width=0.16\textwidth]{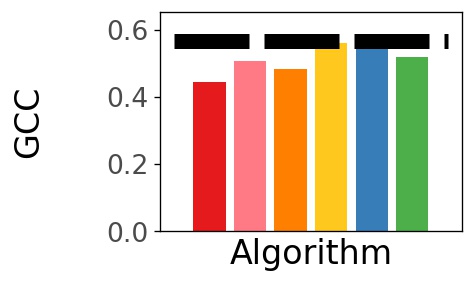}} \\
    \hline
    
    {\multirow{6}{*}{coauth-Geology}} & 
    \raisebox{-.9\totalheight}{\includegraphics[width=0.16\textwidth]{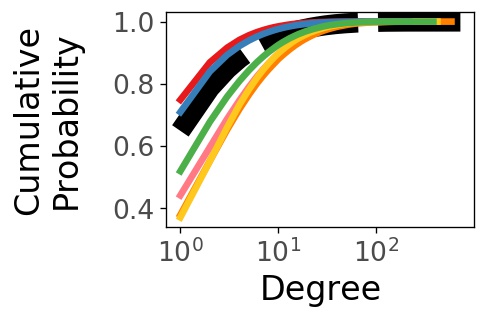}} & 
    \raisebox{-.9\totalheight}{\includegraphics[width=0.16\textwidth]{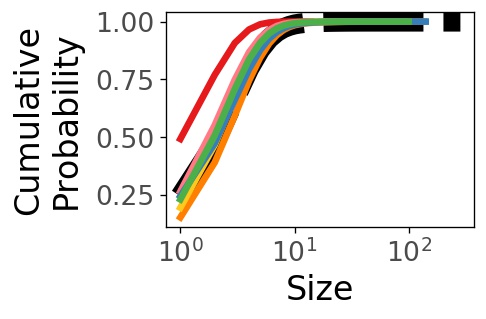}} &
    \raisebox{-.9\totalheight}{\includegraphics[width=0.16\textwidth]{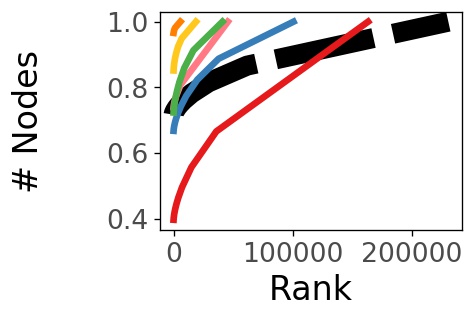}} &
    \raisebox{-.9\totalheight}{\includegraphics[width=0.16\textwidth]{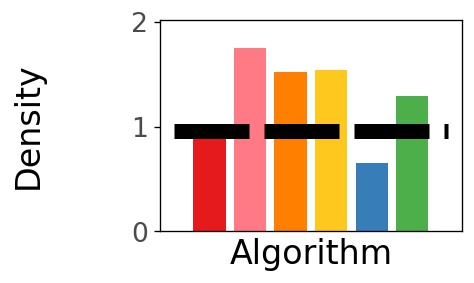}} &
    \raisebox{-.9\totalheight}{\includegraphics[width=0.16\textwidth]{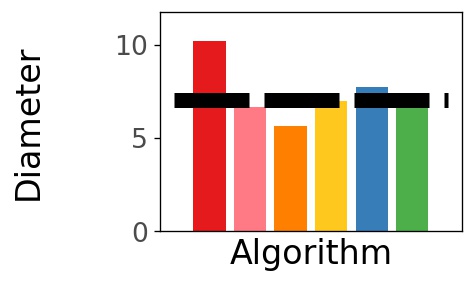}} & 
    \raisebox{-.9\totalheight}{\includegraphics[width=0.16\textwidth]{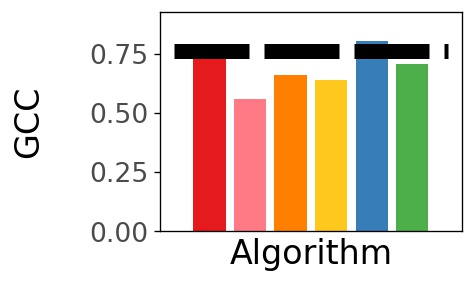}} \\
    
    \bottomrule

 \end{tabular}}
\end{table*}

In this section, we analyze six intuitive approaches for Problem~\ref{problem} with a focus on their empirical characteristics. The findings guide the development of our sampling method \methodauto in Section~\ref{sec:proposed_approach}.

\subsection{Sampling Procedures}

We describe the six intuitive approaches, which are categorized into node-selection methods and hyperedge-selection methods.

\subsubsection{Node Selection (\textsc{NS})} \label{sec:baseline:NS}
In node-selection methods, we choose a subset $\SVH$ of nodes and return the \textit{induced sub-hypergraph} $\SGH=(\SVH,\SEVH)$ where $\SEVH:=\{ e \in \SE: \forall v \in e, v \in \SVH \}$ denotes the set of hyperedges composed only of nodes in $\SVH$.
Each process below is repeated until $\SGH$ has the desired size (i.e., $|\SEVH|=\lfloor |\SE| \cdot p \rfloor$).


\noindent $\circ$ \textbf{Random Node Sampling} (\textit{RNS}): We repeat drawing a node uniformly at random and adding it to $\SVH$.

\noindent $\circ$ \textbf{Random Degree Node} (\textit{RDN}): 
We repeat drawing a node with probabilities proportional to node degrees and adding it to $\SVH$.

\noindent $\circ$ \textbf{Random Walk} (\textit{RW}): We perform random walk with restart \cite{tong2008random} with the restart probability $c=0.15$ on the clique expansion (defined in Section~\ref{problem_form:properties}) and add each visited node to $\SVH$ in turn. We select a new seed node, where random walks restart, after reaching the maximum number of steps, which is set to the number of nodes.

\noindent $\circ$ \textbf{Forest Fire} (\textit{FF}): 
    We simulate forest fire in hypergraphs as in \cite{kook2020evolution}. First, we choose a random node $w$ as an ambassador and burn it. Then, we burn $n$ neighbors of $w$ where $n$ is sampled from a geometric distribution with mean $p/(1-p)$.
    We recursively apply the previous step to each burned neighbor by considering it as a new ambassador, but the number of neighbors to be burned is sampled from a different geometric distribution with mean $q/(1-q)$.
    Each burned node is added to $\SVH$ in turn.
    If there is no new burned node, we choose a new ambassador uniformly at random. 
    We set $p$ to $0.51$ and $q$ to $0.2$ as in \cite{kook2020evolution}.
    This method extends a successful representative sampling method for graphs \cite{leskovec2006sampling} to hypergraphs.

\subsubsection{Hyperedge Selection (HS)}
In hyperedge-selection methods, we draw a subset $\SEH$ of hyperedges and return $\SGH=(\SVEH,\SEH)$, where $\SVEH:=\bigcup_{e\in \SEH}e$ is the set of nodes in any hyperedge in $\SEH$.


\noindent $\circ$ \textbf{Random Hyperedge Sampling} (\textit{RHS}): We draw a target number (i.e., $\lfloor |\SE| \cdot p \rfloor$) of hyperedges uniformly at random.

\noindent $\circ$  \textbf{Totally-Induced Hyperedge Sampling} (\textit{TIHS}): 
We extend totally-induced edge sampling \cite{ahmed2011network} to hypergraphs. We repeat (a) adding a hyperedge uniformly at random to $\SEH$ and (b) adding all hyperedges induced by $\SVEH$ (i.e., $\{ e \in \SE: \forall v \in e, v \in \SVEH \}$) to $\SEH$.


\subsection{Datasets and Evaluation Procedures}
\label{sec:observations:evaluation}

\smallsection{Datasets:}
We use $11$ real-world hypergraphs from six domains~\cite{benson2018simplicial}. We provide details of them in Appendix~\ref{appendix:datasets}

\smallsection{Evaluation:}
In order to compare the qualities of sub-hypergraphs sampled by different methods, we aggregate the ten distances described in Section~\ref{problem_form:problem}.
Since the scales of the distances may differ, we compute rankings and Z-Scores to make it possible to directly compare and average them, as follows: 

\begin{itemize}[leftmargin=*] 
    \item \textbf{Ranking}: With respect to each of \textbf{P1}-\textbf{P10}, we rank all sub-hypergraphs using their distances. 
    \item \textbf{Z-Score}: With respect to each of \textbf{P1}-\textbf{P10}, we standardize the distance of sub-hypergraphs by subtracting the mean and dividing the difference by the standard deviation. 
\end{itemize}

When comparing sampling methods in multiple settings (e.g., target portions and datasets), we compute the above rankings (or Z-Scores) of their samples in each setting and average the rankings (or Z-Scores) of samples from each method.

\subsection{Observations} \label{sec:observations}
We evaluate the six intuitive approaches using the $11$ datasets under five different sampling portions, as described above. The results are summarized in Table~\ref{tab:baseline}. Below, we describe the characteristics of each approach. 
We use $\SGH_{ALG}$ to denote a sub-hypergraph  obtained by each approach $ALG$.

    \subsubsection{Random Node Sampling} \hfill

    \smallsection{Small hyperedges}: In $\GRNS$, large hyperedges are rarely sampled because all nodes in a hyperedge must be sampled for the hyperedge to be sampled, which is unlikely.
    
    \smallsection{Weak connectivity}: As large hyperedges are rare, the local and global connectivity is weak.
    Locally,
    node degrees, node-pair degrees, hyperedge sizes, and intersection sizes tend to be low in $\GRNS$.  Globally, $\GRNS$ tends to have low density, especially low overlapness, and large diameter. It also tends to have many connected components with small portions of nodes.
    
    
    
    \smallsection{Precise preservation of relative singular values}: 
    Relative singular values (see \textbf{P5}) are preserved best in $\GRNS$ among the sub-hypergraphs obtained by node-selection methods.
    
    
    
    
\subsubsection{Random Degree Node, Random Walk, and Forest-Fire} \hfill 

    \smallsection{More high-degree nodes than \textit{RNS}}: 
    \textit{RDS}, \textit{RW}, and \textit{FF} lead to a larger portion of high-degree nodes than \textit{RNS} since they prioritize high-degree nodes. Thus, they preserve degree distributions better than \textit{RNS} in some datasets where \textit{RNS} significantly increases the fraction of low-degree nodes.
    Especially, degree distributions are preserved best by \textit{RDN} in terms of ranking.
    
    
    \smallsection{Stronger connectivity than \textit{RNS}}: 
    High-degree nodes strengthen connectivity.
    Thus, sub-hypergraphs obtained by non-uniform node-selection methods tend to have higher density, higher overlapness, and smaller diameter than $\GRNS$ but
    sometimes even than the original hypergraph $\SG$. Notably, in the sub-hypergraphs, a larger fraction of nodes belong to the largest connected component, reducing the number of connected components, compared to $\SG$.
    

    
    
\subsubsection{Random Hyperedge Sampling} \hfill
    \smallsection{Best preservation of many properties}: 
    \textit{RHS} preserves hyperedge-level statistics (i.e., hyperedge sizes and intersection sizes) near perfectly. 
    It is  also best at preserving connected-component sizes,  relative singular values, and global clustering coefficients.

    
    \smallsection{Weak connectivity}: As \textit{RHS} is equivalent to uniform hypergraph sparsification, $\GRHS$ suffers from weak connectivity. Locally, node degrees and pair degrees tend to be low in $\GRHS$. Globally, in $\GRHS$, density and overlapness are lowest, and diameter is largest. 
    
    

\subsubsection{Totally-Induced Hyperedge Sampling} \hfill
    \smallsection{Complementarity to \textit{RHS}}: \textit{TIHS}  preserves node degrees, node-pair degrees, density, overlapness, and diameter best, which are overlooked by \textit{RHS}. 
    
    \smallsection{Strong connectivity}:
    Still, node degrees, density, and overlapness tend to be higher, and diameter tends to be smaller in $\GTIHS$ than in the original hypergraph $\SG$. That is, 
    the connectivity tends to be a bit stronger in $\GTIHS$ than in $\SG$.
    Thus, $\GTIHS$ tends to have fewer but larger connected components than $\SG$.
    
        
    

\subsubsection{Summary of Observations}
As summarized in Table~\ref{tab:baseline}, when considering all settings, 
\textit{RHS} provides overall the best representative sub-hypergraphs.
While \textit{RHS} produces sub-hypergraphs with weaker connectivity, \textit{RHS} is by far the best method in preserving hyperedge sizes, intersection sizes, relative singular values, connected-component sizes, and global clustering coefficients.

	\section{Proposed Approach}
	\label{sec:proposed_approach}
	\begin{figure}[t]
    \centering
    \vspace{-3mm}
    \begin{minipage}[b] {1.0\linewidth}
    \includegraphics[width=1.\textwidth]{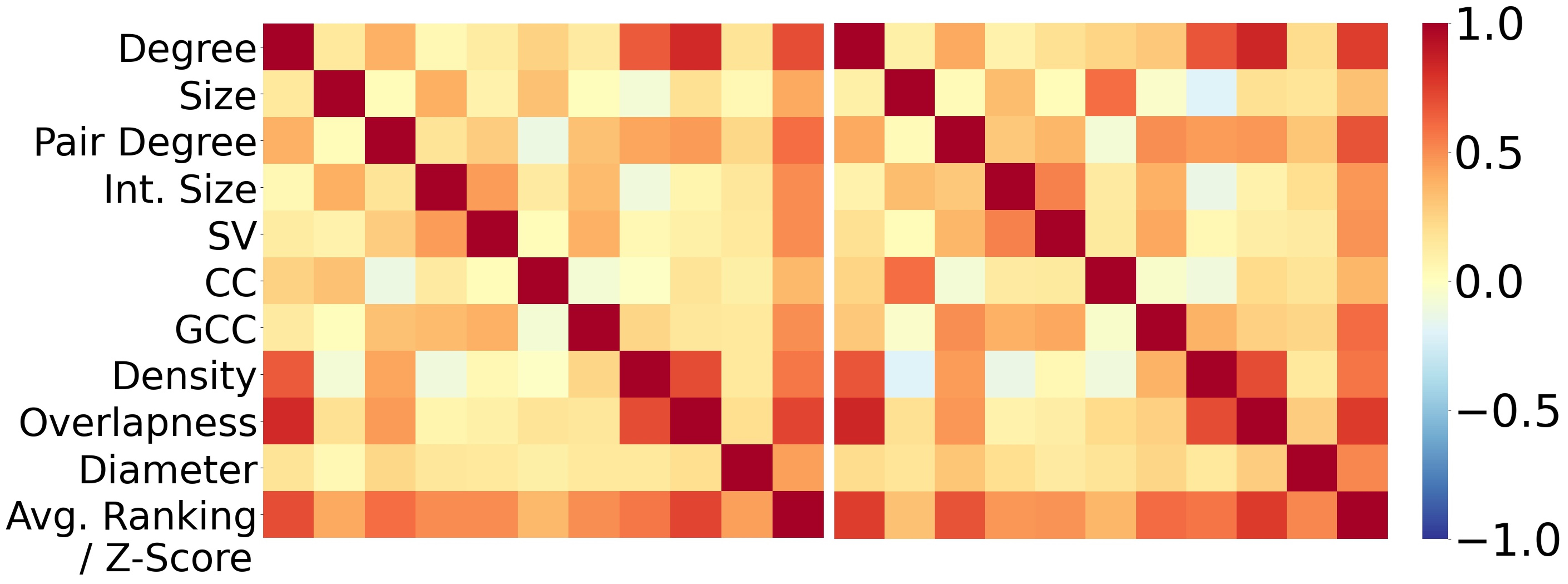}
    \end{minipage}
    
    \vspace{-1mm}
    
    \begin{minipage}{1.0\linewidth} \small
        \hspace{0.27\linewidth}(a) Ranking
        \hspace{0.20\linewidth}(b) Z-Score
    \end{minipage}
    \caption{\label{fig:importance_prop} Pearson correlation coefficients between rankings and Z-scores w.r.t. P1-P10 statistics.
    Overall,
    rankings and Z-scores w.r.t. node degree are most strongly correlated with those w.r.t. the other statistics.
    }
\end{figure}


In this section, we propose \methodauto (\underline{\smash{Mi}}nimum \underline{\smash{D}}egree Bi\underline{\smash{a}}sed \underline{\smash{S}}ampling of Hyperedges), our sampling method for Problem~\ref{problem}.
We first discuss motivations behind it. Then, we describe and analyze \method, a preliminary version of \methodauto.
Lastly, we present the full-fledged version of \methodauto.


\subsection{Intuitions behind \methodauto}
\label{sec:proposed_approach:intuition}

Analyzing the simple approaches in Section~\ref{sec:observations} motivates us to come up with \methodauto. Especially, we focus on the following findings:

\begin{obs}
	\textit{RHS} performs best, but its samples suffer from weak connectivity, including lack of high-degree nodes.
\end{obs} 

\begin{obs}\label{obs:degree}
	The ability to preserve degree distributions is strongly correlated with the abilities to preserve other properties and thus the overall performance, as shown in Figure~\ref{fig:importance_prop}.
\end{obs} 

\begin{figure}[t]
    \vspace{-3mm}
    \centering
    \begin{minipage}[b]{0.37\linewidth}
        \includegraphics[width=1.0\textwidth]{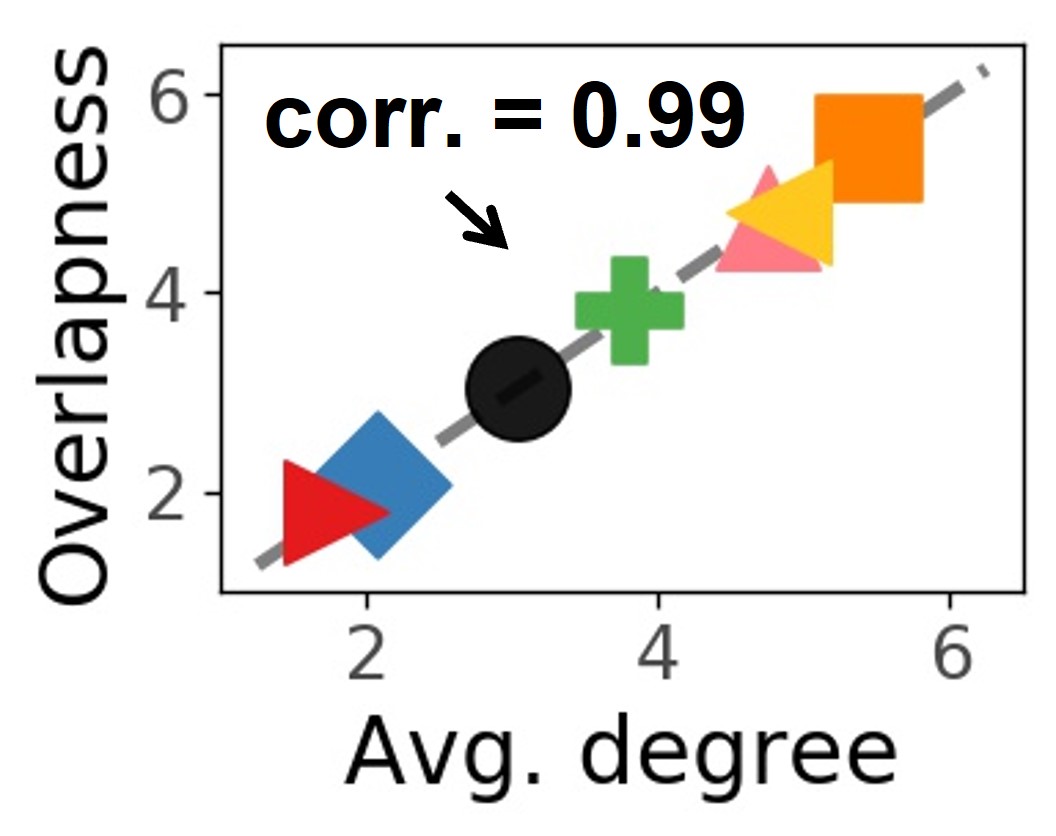}
    \end{minipage}
    \quad
    \begin{minipage}[b]{0.37\linewidth}
        \includegraphics[width=1.0\textwidth]{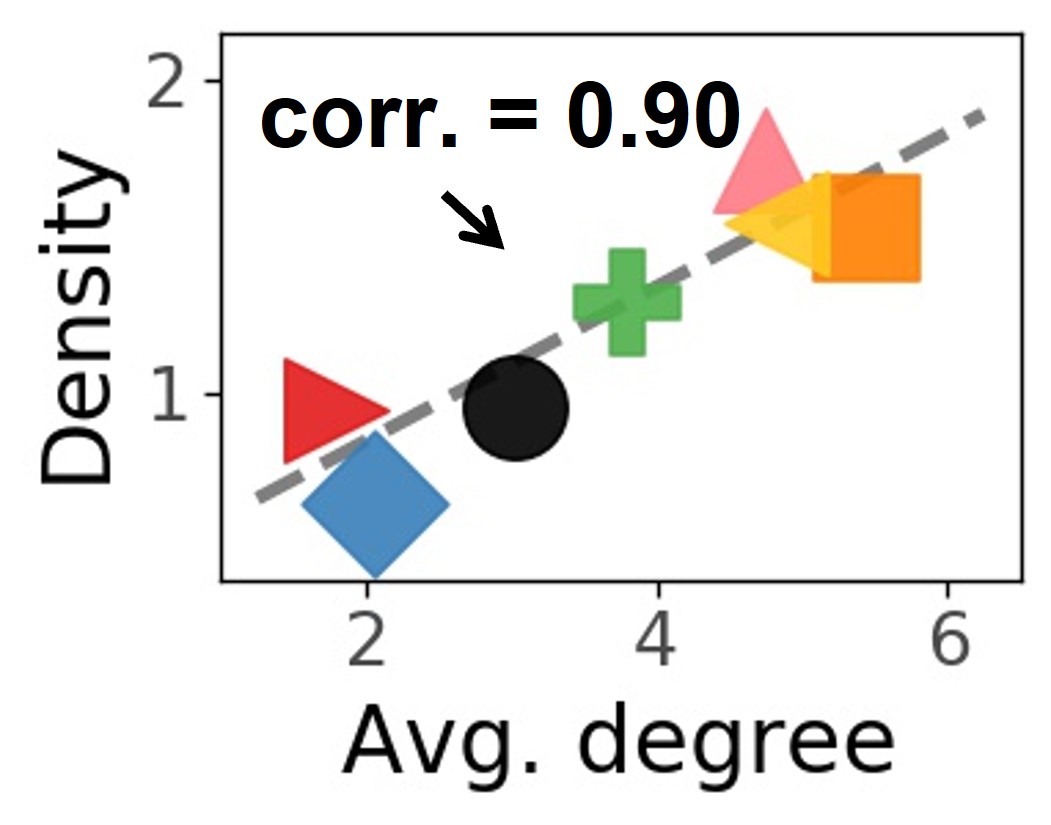}
    \end{minipage}
    \hspace{1mm}
    \begin{minipage}[b] {0.15\linewidth}
        \includegraphics[width=1.0\textwidth]{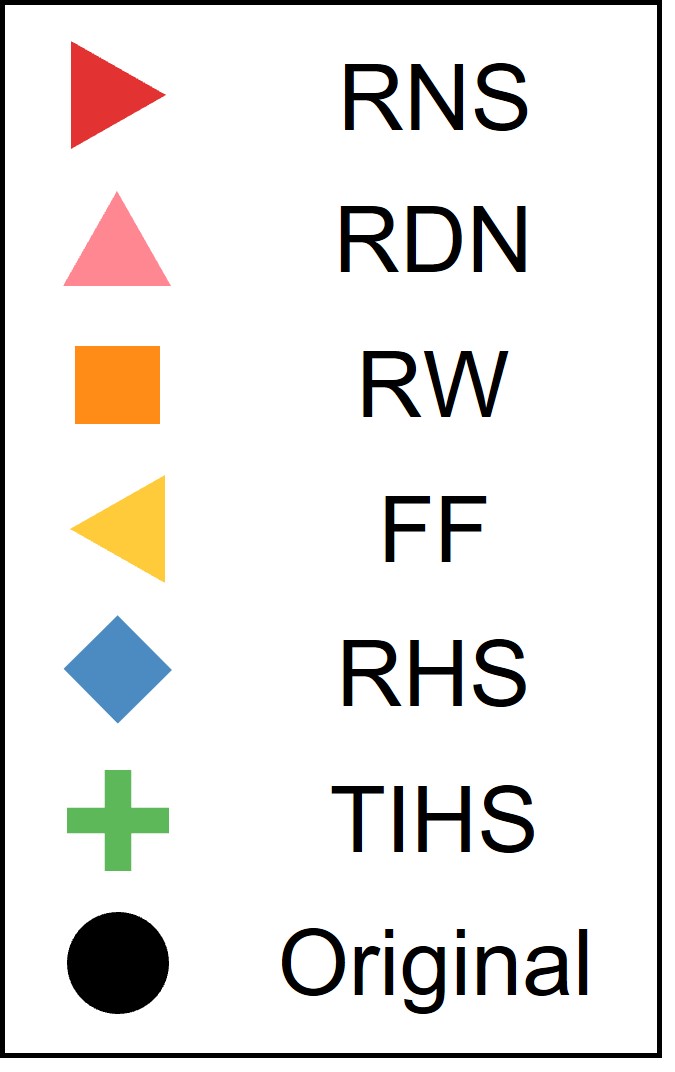}
        \vspace{-0.3mm}
    \end{minipage}
    
    \vspace{-1mm}
    \begin{minipage}{1.0\linewidth} \small
        (a) Overlapness vs. Avg. Degree
        \hspace{0.03\linewidth}(b) Density vs. Avg. Degree
    \end{minipage}
    
    \caption{\label{fig:degree_correlation} Overlapness and density of samples increase as avg. node degree in them increases. We show the results in the coauth-Geology dataset when the sampling ratio is 30\%.
    %
    }
    \vspace{-4mm}
\end{figure}

Specifically, when designing \methodauto, we aim to overcome the limitations of \textit{RHS} while maintaining its strengths.
Especially, based on the above findings, we focus on better preserving node degrees by increasing the fraction of high-degree nodes while expecting that this also helps preserve other properties.
Our expectation is also supported by the strong correlation between (a) the average degree in sub-hypergraphs and (b) their overlapness and density, which tend to be low in sub-hypergraphs sampled by \textit{RHS}.
This correlation, which is shown in Figure~\ref{fig:degree_correlation}, is naturally expected from the fact that high-degree nodes increase the number of hyperedges per node and the definitions of density and overlapness (see Section~\ref{problem_form:properties}). 

\begin{algorithm}[t]
    \small
	\caption{\method: Preliminary Algorithm for Representative Hyperedge Sampling \label{alg:method}}
	\SetKwInOut{Input}{Input}
    \SetKwInOut{Output}{Output}
    \Input{ (1) Hypergraph $\Glong$ \\
    (2) Sampling portion $p\in(0,1)$ \\
    (3) Exponent of hyperedge weight $\alpha\geq 0$}
    \Output{Sub-hypergraph $\Ghat$ of $\Glong$}
    $\SVH \leftarrow \varnothing$ and
    $\SEH \leftarrow \varnothing $ \\
    \While{$|\SEH| < \lfloor |\SE| \cdot p \rfloor $}{
        $\SEH \leftarrow \SEH  \cup\{$a hyperedge $e$ sampled from $\SE \setminus \SEH$ with probability proportional to $\omega(e)^{\alpha}\}$ \hfill \blue{$\triangleright$ $\omega(e)=\min_{v \in e} d_{v}$} \\
    }
    $\SVH \leftarrow \bigcup_{e\in \SEH}e$ \\
    \Return{$\Ghat$}
\end{algorithm}

\subsection{\method: Preliminary Version}
\label{sec:proposed_approach:basic}

How can we better preserve node degrees, which seem to be a decisive property, while maintaining the advantages of \textit{RHS}?
Towards this goal, we first present \method, a preliminary sampling method that determines the amount of bias towards high-degree nodes by a single hyperparameter, for Problem~\ref{problem}.

\subsubsection{Description}
The pseudocode of \method is provided in Algorithm~\ref{alg:method}.
Given a hypergraph $\Glong$ and a sampling portion $p$, it returns a sub-hypergraph $\Ghat$ of $\SG$ where the number of hyperedges in  $\SGH$ is $p$ of that in $\SG$.
Starting from an empty hypergraph, \method repeats drawing a hyperedge as in \textit{RHS}. 
However, unlike \textit{RHS}, 
the probability for each hyperedge $e$ being drawn at each step is proportional to $\omega(e)^{\alpha}$ where $\omega(\cdot)$ is a hyperedge weight function and the exponent $\alpha$ ($\geq 0$) is a given constant.
Note that, if $\alpha$ is zero, \method is equivalent to \textit{RHS}.


Based on the intuitions discussed in Section~\ref{sec:proposed_approach:intuition}, \method prioritizes hyperedges with high-degree nodes to increase the fraction of such nodes.
In order to prioritize especially hyperedges composed \textit{only} of high-degree nodes, it uses $\omega(e):=\min_{v \in e} d_{v}$, where $d_v$ is the degree of $v$ in $\SG$, as the hyperedge weight function.
While $\max_{v \in e} d_{v}$ and  $\mathrm{avg}_{v \in e} d_{v}$ can be considered instead, with them, the degree distribution of $\SG$ cannot be preserved accurately, as shown in Appendix~\ref{appendx:ablation}. 


\subsubsection{Empirical Properties}
The value of $\alpha$ affects the amount of bias towards high-degree nodes in \method.
Below, we analyze how $\alpha$ affects samples (i.e., sub-hypergraphs) in practice. 

\begin{figure}[t]
    \vspace{-3mm}
    \begin{minipage}[b] {0.43\linewidth}
    \includegraphics[width=1.\textwidth]{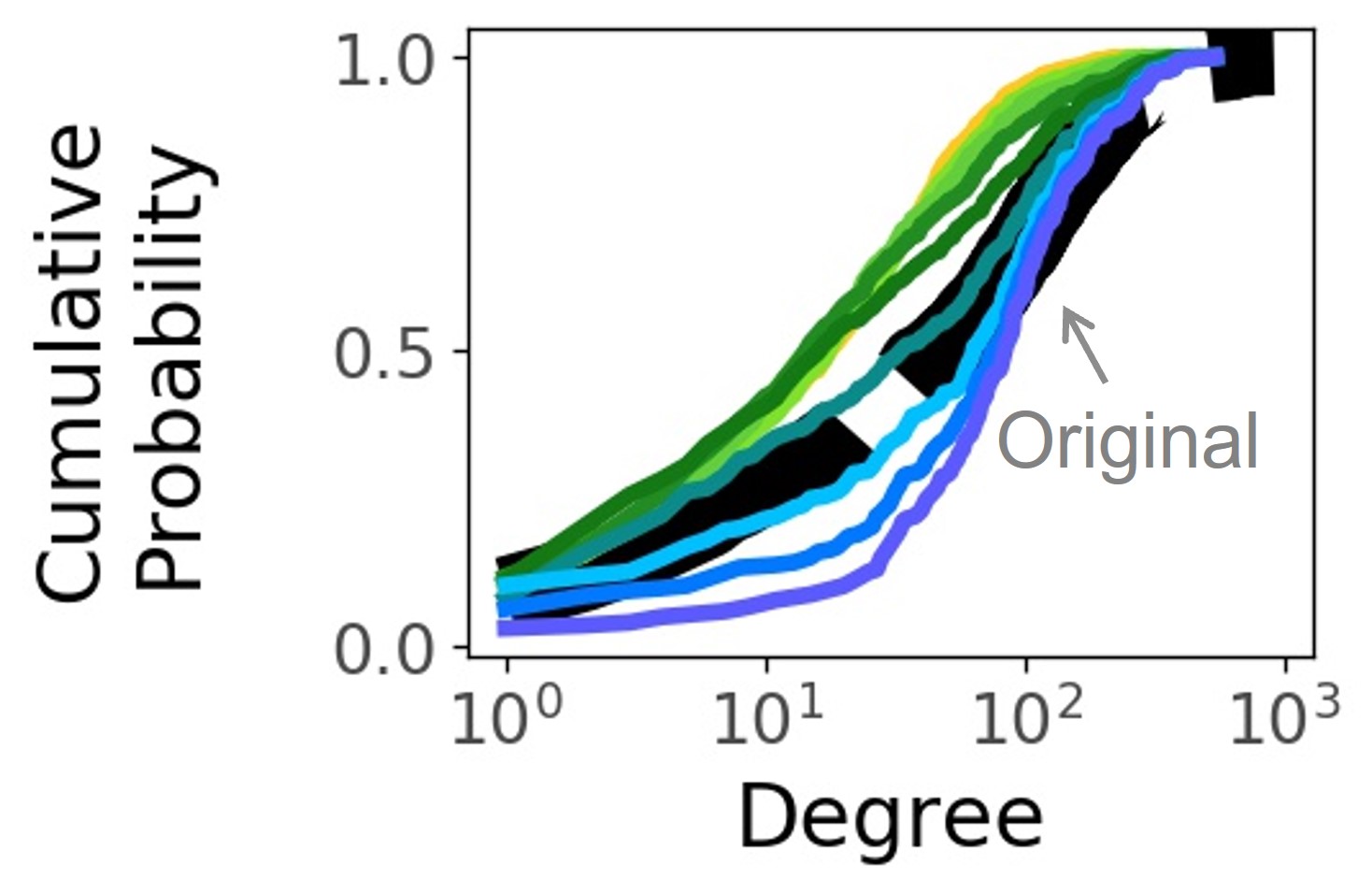}
    \end{minipage}
    \begin{minipage}[b]{0.43\linewidth}
    \includegraphics[width=1.\textwidth]{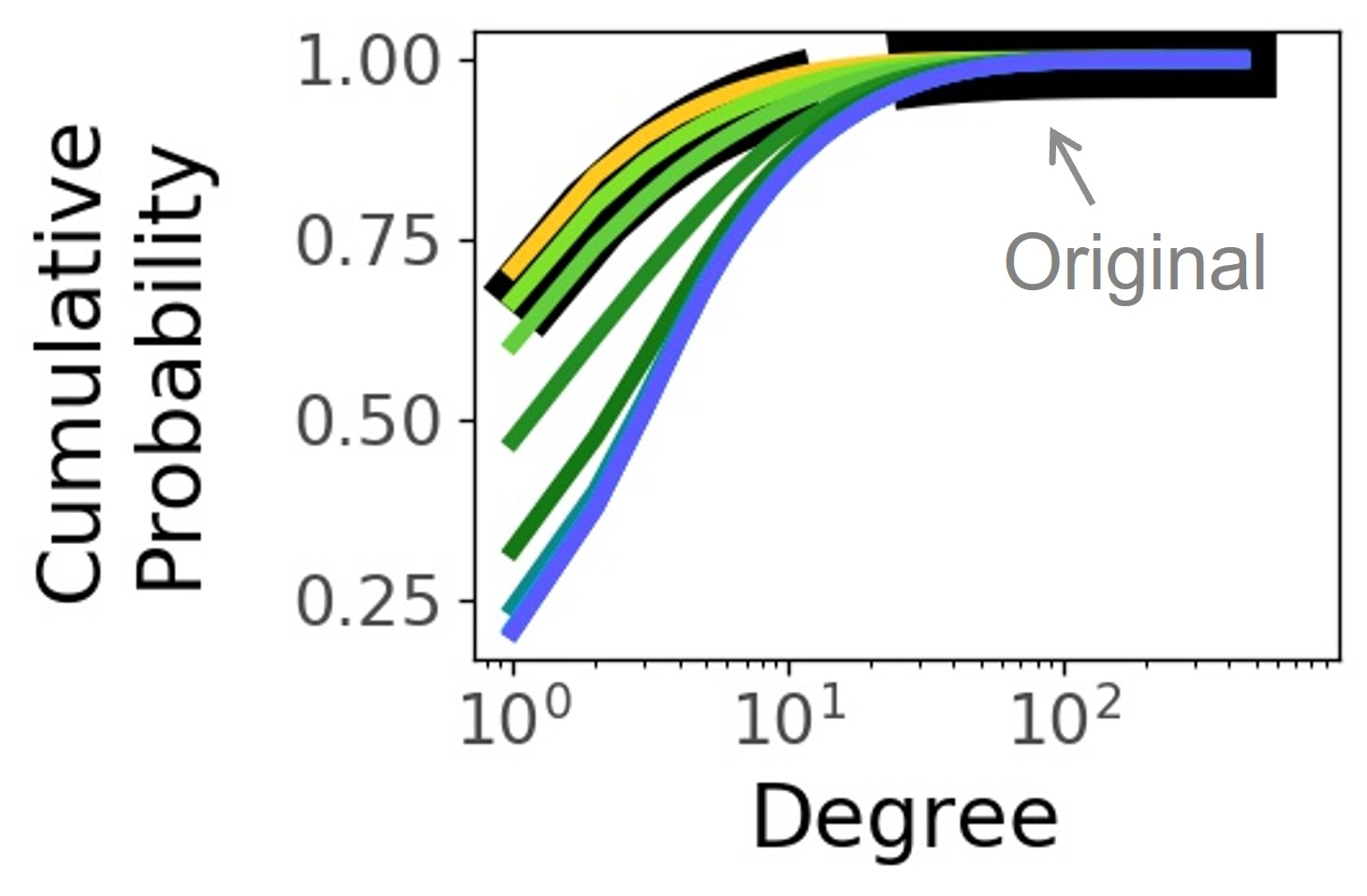}
    \end{minipage}
    \hspace{1mm}
    \begin{minipage}[b]{0.1\linewidth}
    \includegraphics[width=0.9\textwidth]{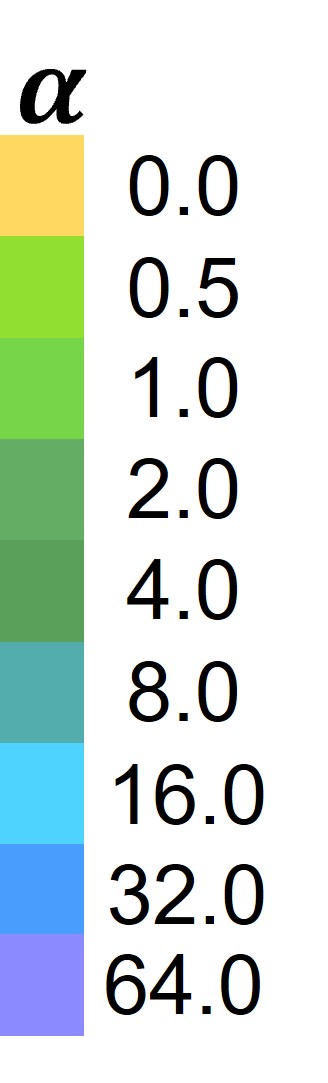}
    \vspace{1mm}
    \end{minipage}
    
    \vspace{-1mm}
    
    \begin{minipage}{1.0\linewidth} \small
        \hspace{0.1\linewidth}(a) email-Eu Dataset
        \hspace{0.17\linewidth}(b) coauth-Geology Dataset
    \end{minipage}
    \caption{Obs~\ref{obs:alpha_and_dist}: Biases in degree distributions in sub-hypergraphs sampled by \method are controlled by $\alpha$. We show the results when the sampling portion is $\bf{0.3}$.}
    \label{fig:alpha_and_dist}
\end{figure}
\begin{figure}[t]
    \vspace{-3mm}
    \centering
    \begin{minipage}[b]{0.36\linewidth}
        \includegraphics[width=1.\textwidth]{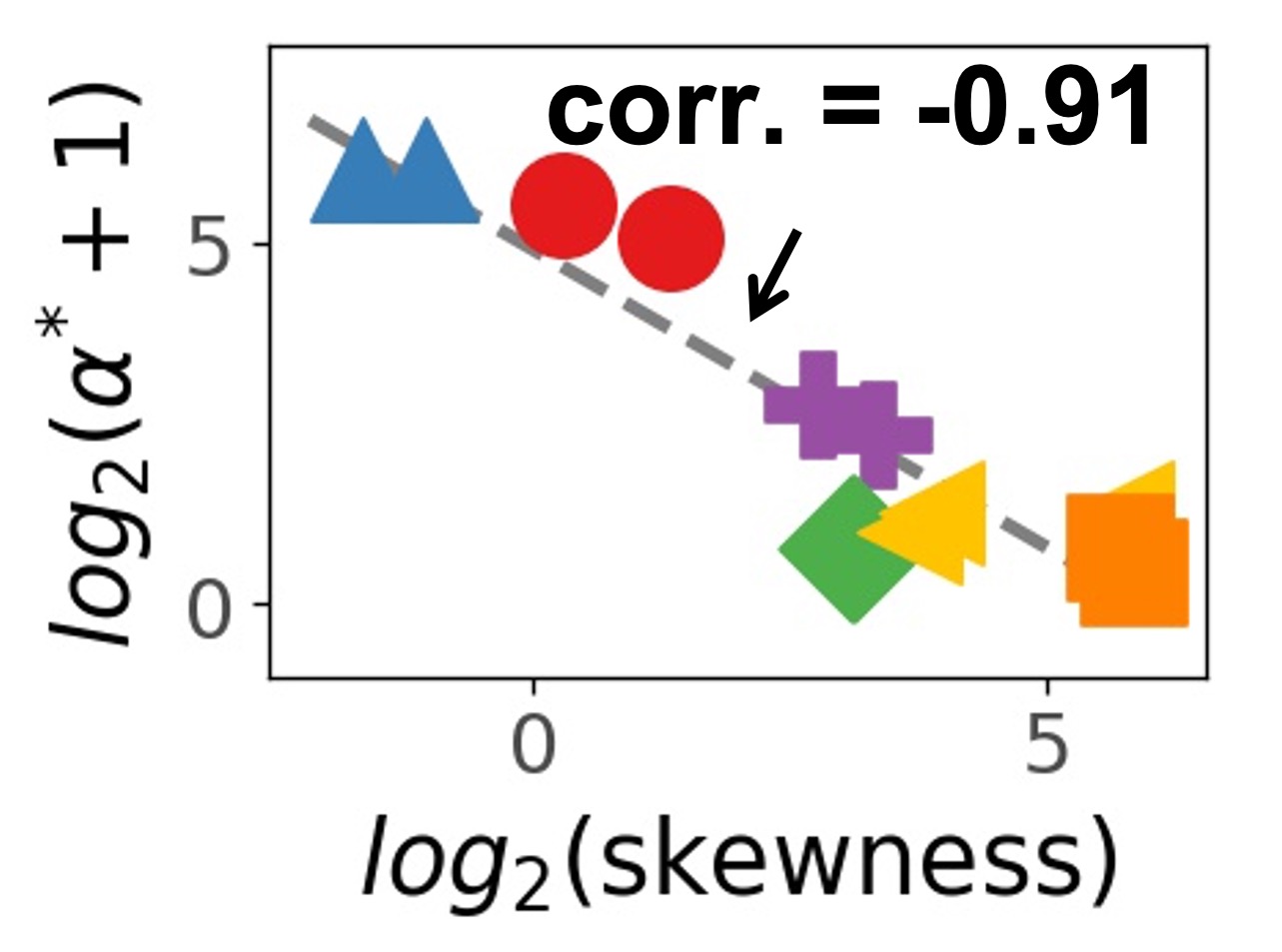}
    \end{minipage}
    \hspace{1mm}
    \begin{minipage}[b]{0.36\linewidth}
        \includegraphics[width=1.\textwidth]{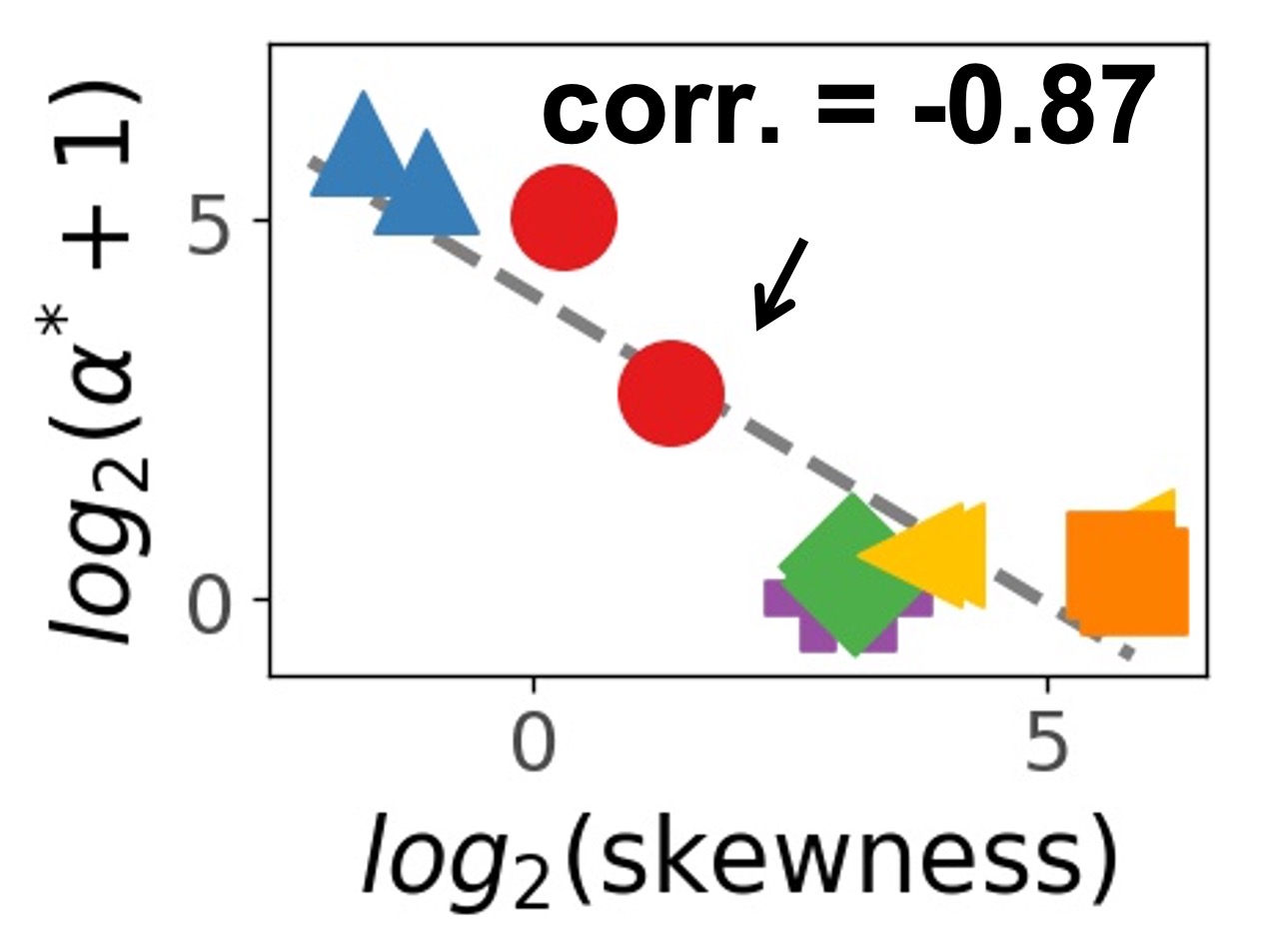}
    \end{minipage}
    \hspace{2mm}
    \begin{minipage}[b]{0.18\linewidth} 
        \includegraphics[width=0.85\textwidth]{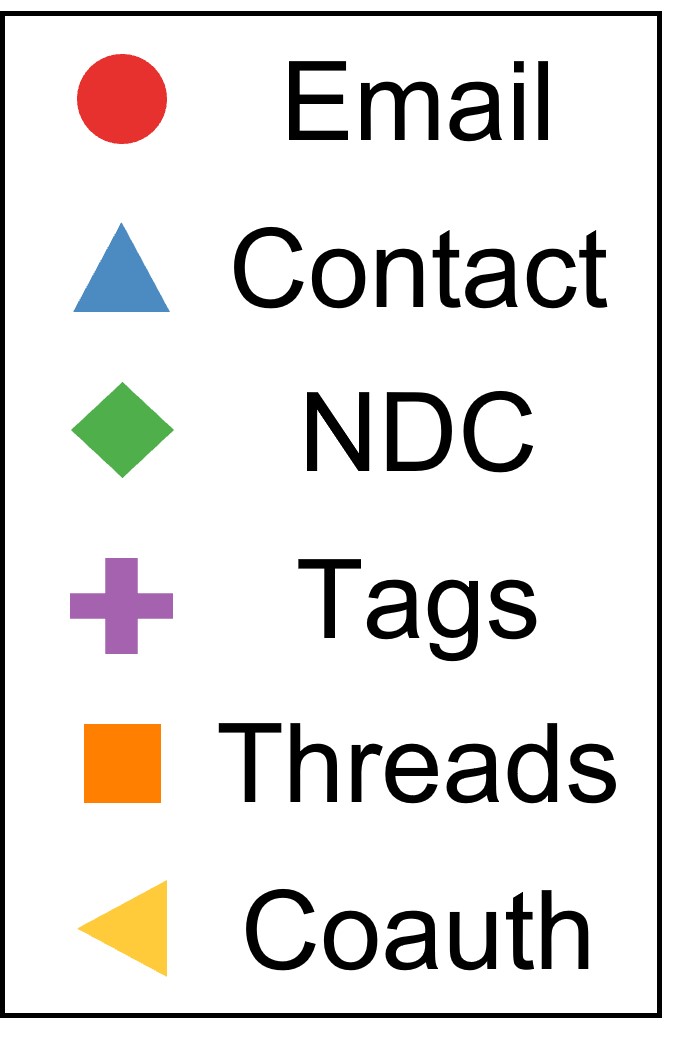}
        \vspace{2mm}
    \end{minipage}
    
    \vspace{-1mm}
    
    \begin{minipage}{1.0\linewidth} \small
        \hspace{0.11\linewidth}{(a) Sampling 10\%}
        \hspace{0.15\linewidth}{(b) Sampling 50\%}
    \end{minipage}
    
     \caption{\label{fig:alpha_and_skewness} Obs.~\ref{obs:alpha_and_skewness}: A strong negative correlation between the skewness of original degree distributions and best performing $\alpha$ values (denoted by $\alpha^*$). Colors denote dataset domains.}
     
\end{figure}

The degrees of nodes within samples obtained with different $\alpha$ values are shown in Figure~\ref{fig:alpha_and_dist}, from which we make Observation~\ref{obs:alpha_and_dist}.
This result is promising, showing that the bias in degree distributions can be directly controlled by $\alpha$.
\begin{obs}\label{obs:alpha_and_dist}
	As $\alpha$ increases, the degree distributions in samples  tend to be more biased towards high-degree nodes.
\end{obs} 

We additionally explore how the best-performing $\alpha$ values,\footnote{They are chosen among $0$, $2^{-3}$, $2^{-2.5}$, $2^{-2}$, $\cdots$ $2^{5.5}$, and $2^{6}$.} which lead to best preservation of degree distributions in terms of D-Statistics (see Section~\ref{problem_form:problem}), are related to the skewness of degree distributions.\footnote{The skewness is defined as $\frac{ \mathrm{E}_{v} [ (d_{v} - \mathrm{E}_{v}[d_{v}])^{3} ]}{ \mathrm{E}_{v}[ (d_{v} - \mathrm{E}_{v} \left[ d_{v} \right] )^{2}]^{3/2} }$.} A strong negative correlation was observed as summarized in Observation~\ref{obs:alpha_and_skewness} and shown in Figure~\ref{fig:alpha_and_skewness}.
\begin{obs}\label{obs:alpha_and_skewness} 
    As degree distributions in original hypergraphs are more skewed, larger $\alpha$ values are required (i.e., high-degree nodes need to be prioritized more) to preserve the distributions.
\end{obs}

\begin{figure}[t]
    \vspace{-3mm}
    \centering
    \begin{minipage}[b]{0.37\linewidth}
    \includegraphics[width=1.0\textwidth]{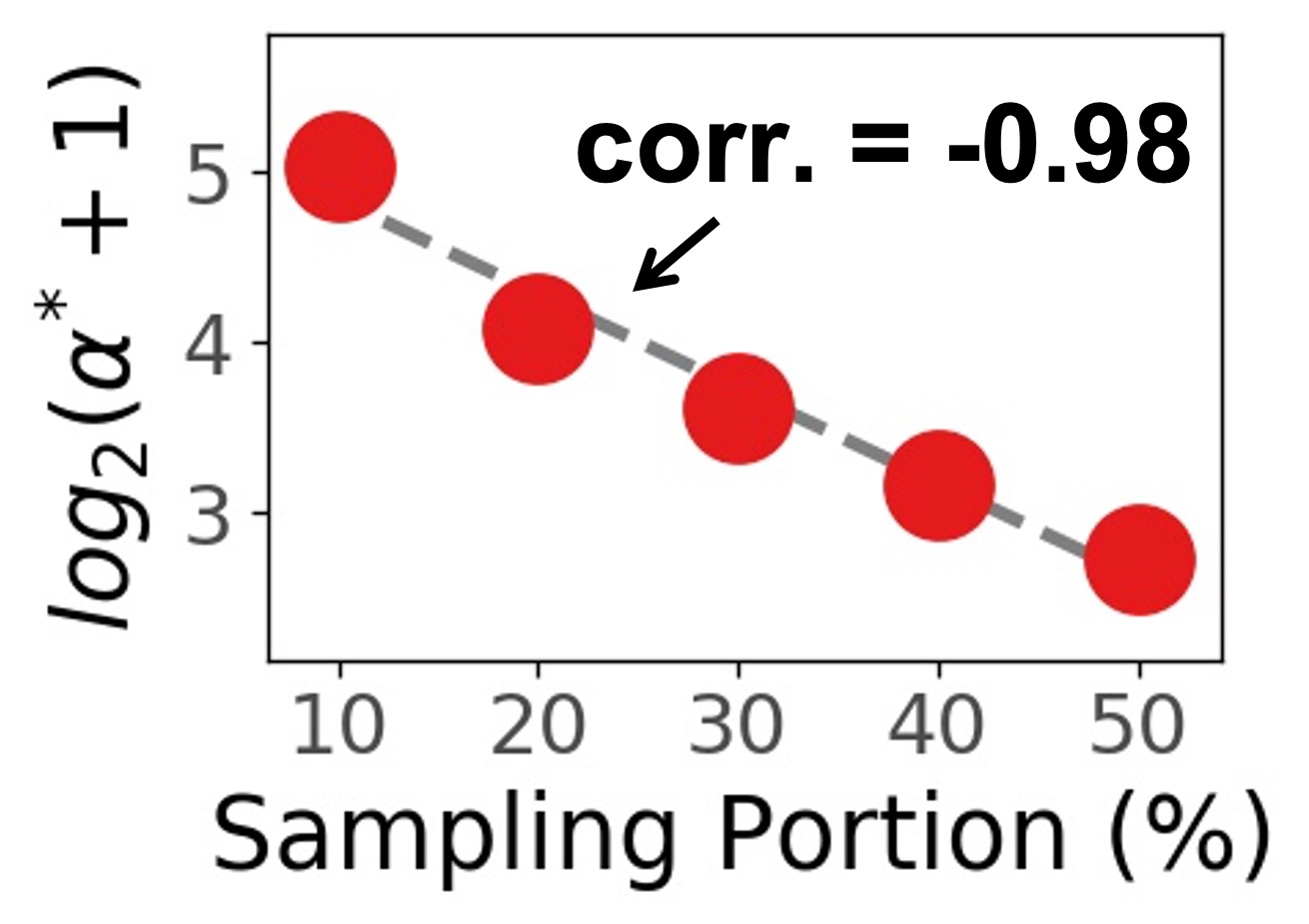}
    \end{minipage}
    \quad
    \begin{minipage}[b]{0.4\linewidth}
    \includegraphics[width=1.0\textwidth]{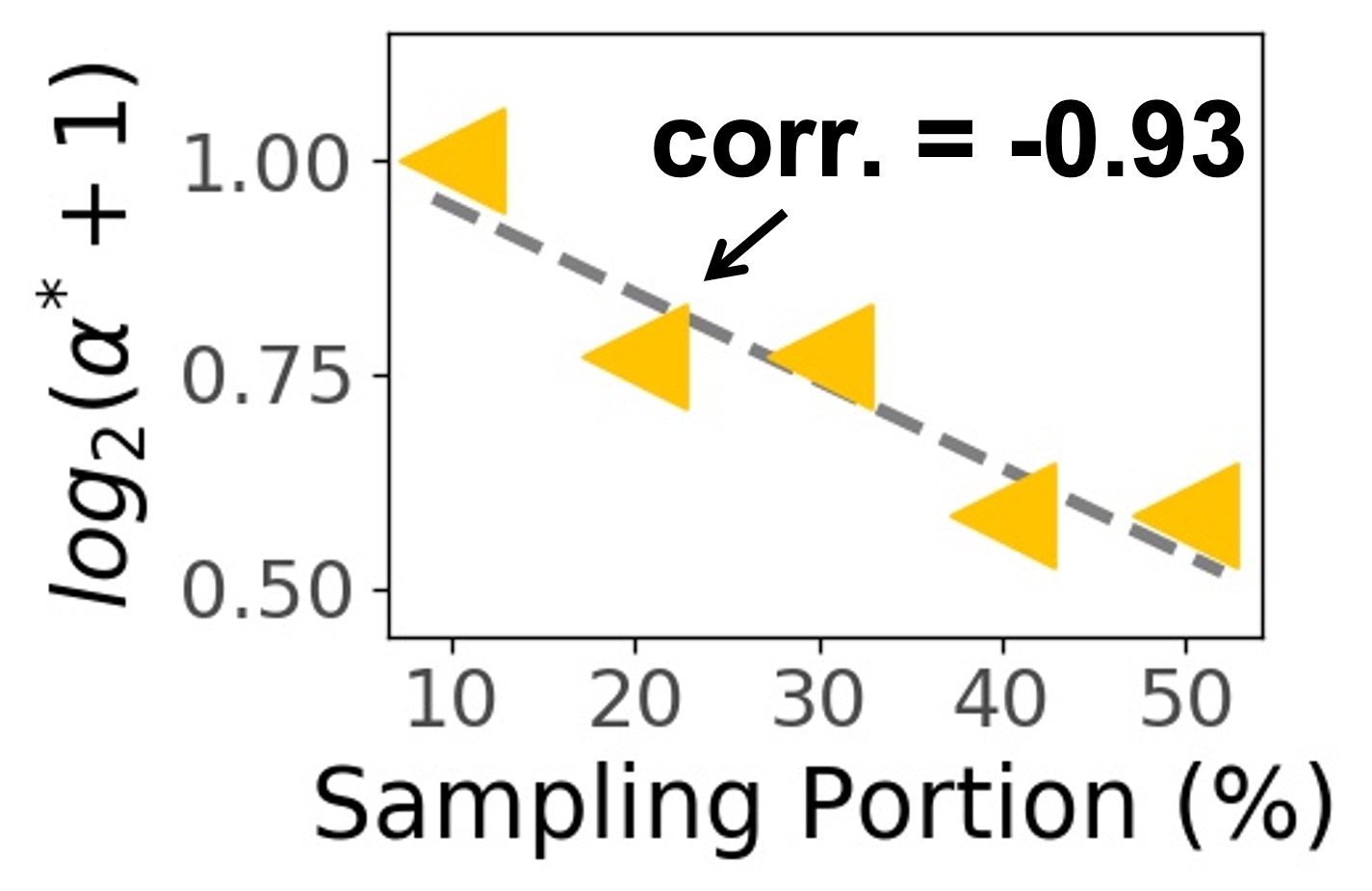}
    \end{minipage}
    
    \vspace{-1mm}
    
    \begin{minipage}{1.0\linewidth} \small
        \hspace{0.16\linewidth}(a) email-Eu Dataset
        \hspace{0.12\linewidth}(b) coauth-Geology Dataset
    \end{minipage}
    \caption{Obs.~\ref{obs:alpha_and_portion}: \label{fig:alpha_and_portion} A strong negative correlation between sampling portions and best-performing $\alpha$ values (denoted by $\alpha^*$).}
\end{figure}

We also find out a strong negative correlation between best-performing $\alpha$ values and sampling portions, as shown in Figure~\ref{fig:alpha_and_portion} and summarized in Observations~\ref{obs:alpha_and_portion}.
\begin{obs}\label{obs:alpha_and_portion} 
	As we sample fewer hyperedges, larger $\alpha$ values are required (i.e., high-degree nodes need to be prioritized more) to preserve degree distributions.
\end{obs}


The observations above are exploited to automatically tune $\alpha$ in the full-fledged version of \method as described in Section~\ref{sec:proposed_approach:full}.


\subsubsection{Theoretical Properties}

We prove the time complexity of \method below.

\begin{thm}[Time Complexity]\label{theorem:time}
The time complexity of Algorithm~\ref{alg:method} is $O(p \cdot |\SE| \cdot \log(\max_{v\in \SV} d_{v})+ \sum_{e\in \SE} |e|)$.
\end{thm}
\begin{proof}
It takes $O(\sum_{e\in \SE} |e|)$ time to compute $\omega(e)$ for every hyperedge $e$, and the other steps take $O(|\SE|+p \cdot |\SE| \cdot \log(\max_{v\in \SV} d_{v}))$ time. See Appendix~\ref{appendix:timecomplexity} for a full proof.
\end{proof}

 In Appendix~\ref{appendix:theoretical_analysis}, we theoretically analyze  Observation~\ref{obs:alpha_and_dist}. Specifically, we provide a condition for bias towards high-degree nodes to grow as $\alpha$ increases,
and we confirm that only $\min_{v \in e} d_{v}$ satisfies this condition, while $\max_{v \in e} d_{v}$ and  $\mathrm{avg}_{v \in e} d_{v}$ do not.



\begin{algorithm}[t]
    \small
	\caption{\methodauto: Proposed Algorithm for Representative Hyperedge Sampling with Automatic $\alpha$ Selection \label{alg:method_auto}}
	\SetKwInOut{Input}{Input}
    \SetKwInOut{Output}{Output}
    \SetKwComment{Comment}{$\triangleright$ }{}
    \SetKwFunction{algo}{algo}\SetKwFunction{proc}{initialize}
    \Input{(1) Hypergraph $\Glong$ \\ (2) Sampling portion $p$\\(3) Trained linear regression model $\SM$\\(3) Search space $\SSS$ for $\alpha$}
    \Output{Sub-hypergraph $\Ghat$ of $\Glong$}
    
    $s \leftarrow $ skewness of the degree distribution in $\SG$ \\
    $\alpha' \leftarrow \SM (s, p)$ \hfill \blue{$\triangleright$ Search based on Observations~\ref{obs:alpha_and_skewness} and \ref{obs:alpha_and_portion}}\\
    $\alpha^{*} \leftarrow$ hill-climbing search within $\SSS$ from $\alpha'$  to minimize $\mathcal{L'}(\alpha)$ \\
    \nonl \nonumber \ \hfill \blue{$\triangleright$ $\mathcal{L'}(\alpha):=\mathcal{L}(\SG,\method(\SG, p, \alpha))$}\\
    \Return{$\method(\SG, p, \alpha^{*})$}
\end{algorithm}

\subsection{\methodauto: Full-Fledged Version}
\label{sec:proposed_approach:full}

As suggested by Observation~\ref{obs:alpha_and_dist}, the hyperparameter $\alpha$ in \method should be tuned carefully. We propose \methodauto, the full-fledged version of our sampling method that automatically tunes $\alpha$.

Based on the strong correlations in Observations \ref{obs:alpha_and_skewness} and \ref{obs:alpha_and_portion}, \methodauto tunes $\alpha$ using a linear regressor $\SM$ that maps (a) the skewness of the degree distribution in the input hypergraph $\SG$ and (b) the sampling portion to (c) a best-performing $\alpha$ value.
In our experiments in Section~\ref{sec:evaluation}, $\SM$ was fitted using the best-performing $\alpha$ values\footnote{We chose $\alpha$ minimizing  $\mathcal{L}(\SG, \SGH)$ among $\mathcal{S}=\{0, 2^{-3}, 2^{-2.5}, \cdots, 2^{5.5}, 2^6\}$.} on the considered datasets with five different sampling portions.\footnote{We set $p$ to $10\%$, $20\%$, $30\%$, $40\%$, or $50\%.$} For a fair comparison, when evaluating \methodauto on a dataset, we used only the remaining datasets for fitting $\SM$.

The $\alpha$ value obtained by the linear regression model $\SM$ is further tuned using hill climbing \cite{russell2002artificial}.
As the objective function, $\mathcal{L}(\SG, \SGH)$,
\methodauto uses the \textit{D-statistics} (see Section~\ref{problem_form:problem}) between the degree distributions in the input hypergraph $\SG$ and a sample $\SGH$.
For speed, we search for $\alpha$ within a given discrete search space $\SSS$,\footnote{We set $\mathcal{S}=\{0, 2^{-3}, 2^{-2.5}, \cdots, 2^{5.5}, 2^6\}$ throughout the paper.} aiming to minimize $\mathcal{L'}(\alpha):=\mathcal{L}(\SG,\method(\SG, p, \alpha))$.
A search ends when it  (a) finds a local minimum of $\alpha$ when we limit our attention to $\mathcal{S}$ or (b) reaches an end of $\mathcal{S}$. Algorithm~\ref{alg:method_auto} describes \methodauto.

The worst-case time complexity of \methodauto is $O(|\SSS|\cdot |\SE| + p \cdot |\SSS|\cdot |\SE| \cdot \log(\max_{v\in \SV} d_{v})+\sum_{e\in \SE} |e|)$. Computing the skewness $s$ and $\omega(e)$ for every hyperedge $e$ takes $O(\sum_{e\in \SE} |e|)$ time. Once they are given, \method is executed at most $|\SSS|$ times.
However, \methodauto rarely searches $\SSS$ exhaustively
as shown empirically in Section~\ref{sec:evaluation:speed}. 
Since we expect users to simply apply new hypergraphs to trained $\SM$,\footnote{$\SM(s, p) = -2.358277 \cdot s - 0.849945 \cdot p + 5.261775$, when it is fitted to the $11$ considered real-world hypergraphs.}
 we do not consider the time required for (re-)training $\SM$.

\begin{table*}[t!]
    \vspace{-3mm}
\caption{\label{tab:evaluation2} 
\methodauto yields overall the most representative sub-hypergraphs. We compare 13 sampling methods on $\bf{11}$ real-world hypergraphs with $\bf{5}$ different sampling portions. We report their distances (D-statistics or differences), Z-Scores, and rankings, as described in Sections~\ref{problem_form:problem} and \ref{sec:observations:evaluation}. The smaller the measures are, the more representative the samples are.}
    \scalebox{0.82}{
    \renewcommand{\arraystretch}{0.8}
        \begin{tabular}{c | c| rr r r r r rr | rr r rr r r rr || rrr}
            \toprule
            \multicolumn{2}{c|}{} && \multirow{2}{*}{\textit{RNS}} & \multirow{2}{*}{\textit{DNS}} & \multirow{2}{*}{\textit{RW}} & \multirow{2}{*}{\textit{FF}} & \multirow{2}{*}{\textit{RHS}} & \multirow{2}{*}{\textit{TIHS}} &&& \multicolumn{3}{c}{\textbf{MGS - Deg}} && \multicolumn{3}{c}{\textbf{MGS - Avg}} &&& \multirow{2}{*}{\textbf{\methodauto}} & \\
            \multicolumn{2}{c|}{} && &  &  &  & &  &&& Add & Rep & Del && Add & Rep & Del &&&  & \\
            \midrule
            \midrule
            \multirow{3}{*}{\textbf{Degree}} &  Dstat. && 0.291 & 0.285 & 0.317 & 0.302 & 0.302 & 0.283 &&& 0.241 & 0.257 & 0.217 && 0.285 & 0.270 & 0.259 &&& \textbf{0.133} & \\
             & Rank && 8.273 & 7.400 & 8.800 & 8.873 & 10.018 & 7.509 &&& 5.509 & 5.945 & 3.727 && 8.473 & 7.764 & 6.018 &&& \textbf{2.691} & \\
             &  Z-Score && 0.241 & 0.183 & 0.693 & 0.573 & 0.240 & 0.179 &&& -0.307 & -0.114 & -0.479 && 0.098 & -0.013 & -0.091 &&& \textbf{-1.203} & \\
            \midrule
            \multirow{3}{*}{\textbf{Int. Size}} & Dstat. && 0.093 & 0.033 & 0.038 & 0.035 & 0.007 & 0.033 &&& 0.014 & 0.024 & 0.053 && 0.002 & \textbf{0.002} & 0.008 &&& 0.024 & \\
             & Rank && 10.145 & 9.109 & 9.473 & 10.364 & 3.491 & 9.473 &&& 4.764 & 5.909 & 8.691 && \textbf{2.545} & 3.200 & 5.036 &&& 8.782 & \\
             &  Z-Score && 1.170 & 0.263 & 0.466 & 0.709 & -0.717 & 0.294 &&& -0.527 & -0.432 & 0.410 && \textbf{-0.776} & -0.734 & -0.424 &&& 0.298 & \\
            \midrule
            \multirow{3}{*}{\textbf{Pair Degree}} &  Dstat. && 0.132 & 0.111 & 0.089 & 0.112 & 0.112 & 0.090 &&& 0.092 & 0.089 & 0.064 && 0.089 & 0.075 & \textbf{0.063} &&& 0.094 & \\
             &  Rank && 8.145 & 9.055 & 6.636 & 8.909 & 9.582 & 6.382 &&& 7.309 & 7.636 & 5.673 && 6.164 & 5.600 & \textbf{4.036} &&& 5.873 & \\
             &  Z-Score && 0.546 & 0.588 & -0.069 & 0.440 & 0.312 & -0.215 &&& -0.002 & -0.001 & -0.409 && -0.058 & -0.239 & \textbf{-0.492} &&& -0.402 & \\
            \midrule
            \multirow{3}{*}{\textbf{Size}} &  Dstat. && 0.227 & 0.105 & 0.121 & 0.057 & 0.009 & 0.085 &&& 0.020 & 0.034 & 0.099 && 0.007 & \textbf{0.003} & 0.023 &&& 0.051 & \\
             &  Rank && 12.655 & 10.691 & 10.400 & 8.164 & 3.582 & 9.691 &&& 4.927 & 5.418 & 8.727 && 2.418 & \textbf{1.673} & 4.600 &&& 8.055 & \\
             &  Z-Score && 2.177 & 0.520 & 0.708 & -0.135 & -0.723 & 0.280 &&& -0.611 & -0.465 & 0.383 && -0.761 & \textbf{-0.794} & -0.432 &&& -0.148 & \\
            \midrule
            \multirow{3}{*}{\textbf{SV}} &  Dstat. && 0.122 & 0.158 & 0.154 & 0.164 & \textbf{0.084} & 0.154 &&& 0.101 & 0.087 & 0.115 && 0.085 & 0.085 & 0.096 &&& 0.125 & \\
             &  Rank && 6.945 & 9.218 & 8.509 & 10.309 & \textbf{3.600} & 8.691 &&& 5.364 & 4.473 & 6.436 && 4.273 & 4.036 & 4.745 &&& 7.309  & \\
             &  Z-Score && 0.215 & 0.760 & 0.523 & 0.839 & \textbf{-0.650} & 0.500 &&& -0.432 & -0.520 & 0.141 && -0.484 & -0.565 & -0.356 &&& 0.028 & \\
            \midrule
            \multirow{3}{*}{\textbf{CC}} &  Dstat. && 0.160 & 0.132 & 0.175 & 0.142 & \textbf{0.097} & 0.135 &&& 0.104 & 0.102 & 0.103 && 0.101 & 0.100 & 0.101 &&& 0.115 & \\
             &  Rank && 8.273 & 6.673 & 8.564 & 7.636 & 3.582 & 7.018 &&& 4.964 & 4.618 & 5.382 && 3.891 & \textbf{3.182} & 4.745 &&& 6.473 & \\
             &  Z-Score && 1.117 & 0.172 & 0.925 & 0.334 & -0.441 & 0.185 &&& -0.338 & -0.444 & -0.314 && -0.334 & \textbf{-0.492} & -0.350 &&& -0.021 & \\
            \midrule
            \multirow{3}{*}{\textbf{GCC}} &  Diff. && 0.120 & 0.153 & 0.085 & 0.119 & 0.100 & 0.083 &&& 0.097 & 0.096 & 0.093 && 0.106 & 0.099 & \textbf{0.075} &&& 0.081 & \\
             &  Rank && 7.509 & 9.691 & 7.055 & 8.636 & 6.818 & 6.909 &&& \textbf{5.309} & 6.309 & 7.491 && 6.309 & 6.436 & 6.473 &&& 6.055 & \\
             &  Z-Score && 0.448 & 0.870 & -0.076 & 0.348 & -0.152 & -0.051 &&& -0.281 & -0.244 & -0.029 && -0.212 & -0.249 & \textbf{-0.307} &&& -0.064 & \\
            \midrule
            \multirow{3}{*}{\textbf{Density}} &  Diff. && 0.374 & 0.540 & 0.488 & 0.516 & 0.523 & 0.426 &&& 0.400 & 0.500 & 0.520 && 0.492 & 0.501 & 0.509 &&& \textbf{0.202} & \\
             &  Rank && 4.636 & 8.436 & 6.655 & 7.164 & 9.727 & 6.127 &&& 5.164 & 7.545 & 8.545 && 6.818 & 7.800 & 8.782 &&& \textbf{2.236} & \\
             &  Z-Score && -0.412 & 0.501 & -0.010 & 0.078 & 0.371 & -0.316 &&& -0.209 & 0.240 & 0.367 && 0.180 & 0.246 & 0.295 &&& \textbf{-1.333} & \\
            \midrule
            \multirow{3}{*}{\textbf{Overlapness}} &  Diff. && 0.546 & 0.550 & 0.616 & 0.531 & 0.523 & 0.460 &&& 0.404 & 0.472 & 0.451 && 0.501 & 0.500 & 0.489 &&& \textbf{0.202} & \\
             &  Rank && 8.673 & 6.891 & 7.782 & 7.400 & 10.182 & 5.764 &&& 5.491 & 7.018 & 5.600 && 8.309 & 8.418 & 7.327 &&& \textbf{2.145} & \\
             &  Z-Score && 0.351 & 0.006 & 0.516 & -0.064 & 0.300 & -0.321 &&& -0.132 & 0.095 & -0.008 && 0.213 & 0.201 & 0.162 &&& \textbf{-1.318} & \\
            \midrule
            \multirow{3}{*}{\textbf{Diameter}} &  Diff. && 0.344 & 0.139 & 0.117 & 0.109 & 0.195 & 0.117 &&& 0.157 & 0.162 & 0.158 && 0.204 & 0.182 & 0.182 &&& \textbf{0.079} & \\
             &  Rank && 9.400 & 6.364 & 7.145 & 5.855 & 8.036 & 5.636 &&& 5.364 & 7.400 & 7.345 && 8.200 & 8.345 & 8.182 &&& \textbf{3.727} & \\
             &  Z-Score && 1.106 & -0.140 & -0.071 & -0.296 & 0.223 & -0.364 &&& -0.322 & -0.009 & 0.016 && 0.282 & 0.191 & 0.177 &&& \textbf{-0.795} & \\
            \midrule
            \midrule
            \multirow{2}{*}{\textbf{Average}} &  Rank && 8.465 & 8.353 & 8.102 & 8.331 & 6.862 & 7.320 &&& 5.416 & 6.227 & 6.762 && 5.740 & 5.645 & 5.995 &&& \textbf{5.335} & \\
             & Z-Score && 0.696 & 0.372 & 0.360 & 0.283 & -0.123 & 0.017 &&& -0.316 & -0.189 & 0.008 && -0.185 & -0.245 & -0.182 &&& \textbf{-0.496} & \\
            \bottomrule
        \end{tabular}}
\end{table*}

	\section{Evaluation}
	\label{sec:evaluation}
	\begin{figure*}[!t]
    \vspace{-3mm}
    \centering
    \begin{minipage}[c]{0.83\linewidth}
        \includegraphics[width=\textwidth]{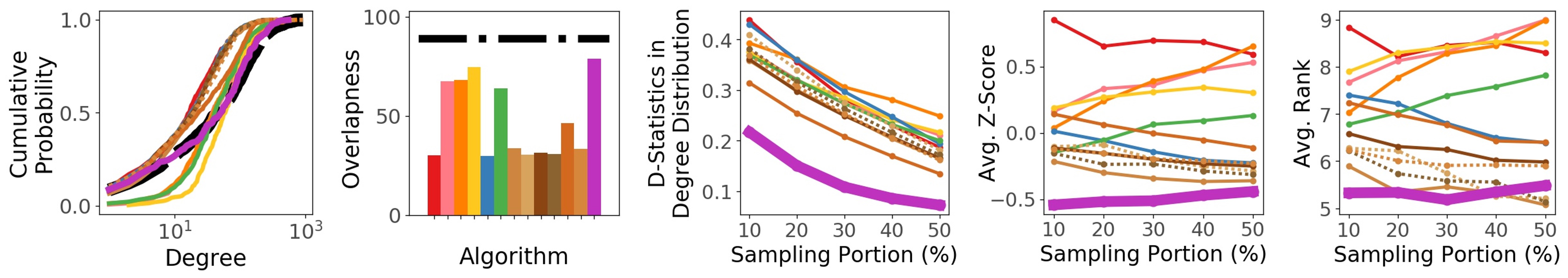}
    \end{minipage}
    \begin{minipage}[c]{0.15\linewidth} 
        \includegraphics[width=1.0\textwidth,]{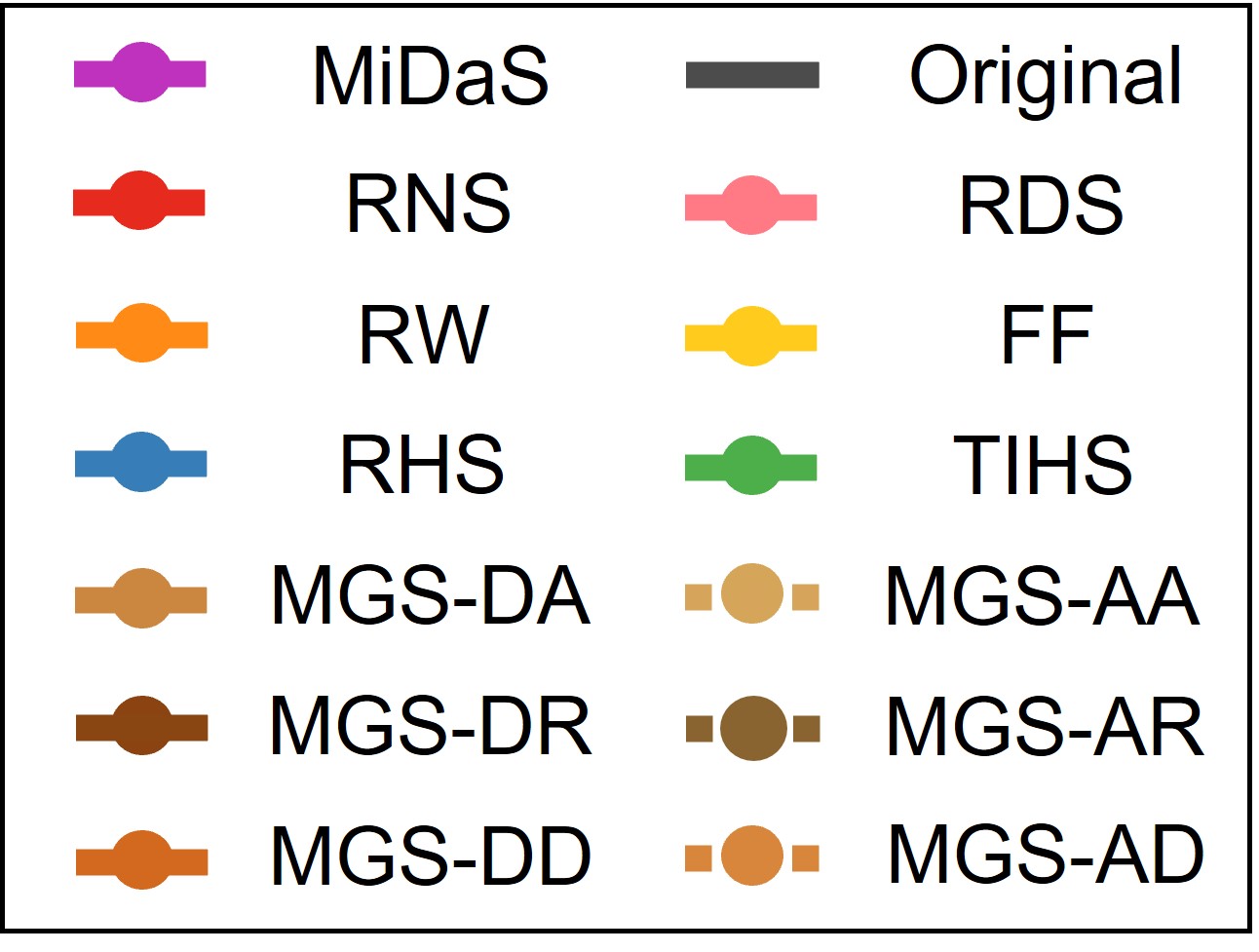}
        \vspace{1mm}
    \end{minipage}
    
    \begin{minipage}{1.0\linewidth} \small
        \hspace{0.03\linewidth}(a) Degree Distribution
        \hspace{0.04\linewidth}(b) Overlapness
        \hspace{0.04\linewidth}(c) Degree Preservation
        \hspace{0.042\linewidth}(d) Avg. Z-Score
        \hspace{0.045\linewidth}(e) Avg. Ranking
    \end{minipage}
    
    \caption{(a-b) \methodauto (purple solid lines) preserves the degree distribution and overlapness of the original hypergraph (black dashed lines) best when sampling 30\% of hyperedges in the email-Eu dataset. (c-e) \methodauto consistently performs best regardless of sampling portions with few exceptions. \label{fig:evaluation:all}
    }
\end{figure*}

\begin{figure}[t]
    \centering
    \begin{minipage}{0.75\linewidth}
        \includegraphics[width=1.0\textwidth]{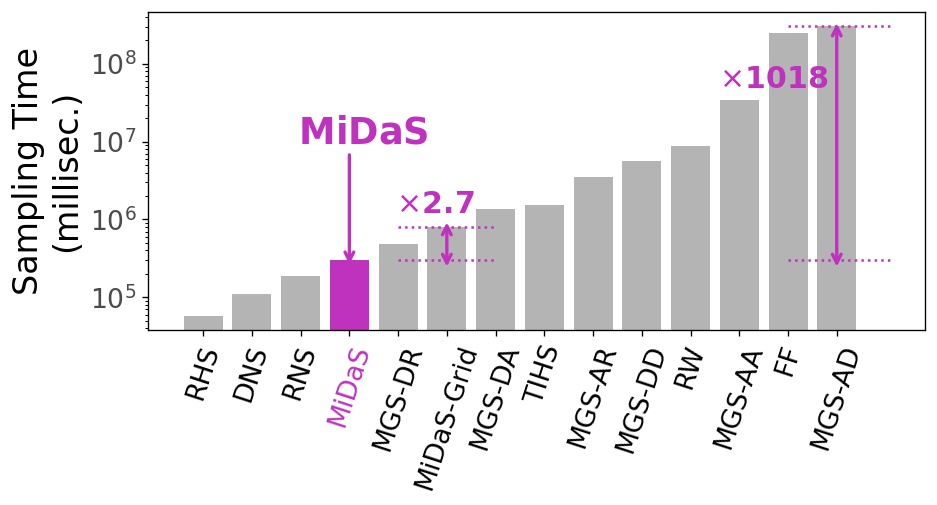}
    \end{minipage}
    \caption{Except for the simplest ones (\textit{RHS}, \textit{DNS}, and \textit{RNS}), \methodauto is the fastest among the considered sampling methods.
    The automatic hyperparameter tuning by \methodauto is $\bf{2.7}\times$ faster than grid search (see \methodgrid).
    }
    \label{fig:evaluation4:time}
\end{figure}

We design experiments to answer the following questions:
\begin{enumerate}[leftmargin=*, label=\textbf{Q\arabic*.}]
    \item \textbf{Quality:} How well does \methodauto preserve the ten structural properties (\textbf{P1}-\textbf{P10}) of real-world hypergraphs? 
    \item \textbf{Robustness:} Does \methodauto perform well regardless of the sampling portions? 
    \item \textbf{Speed:} How fast is \methodauto compared to the competitors?
\end{enumerate}


\subsection{Experimental Settings}
\label{sec:evaluation:settings}

We used the $11$ datasets described in Appendix~\ref{appendix:datasets}. 
We compared \methodauto with the simple methods described in Section~\ref{sec:baseline} and six versions of metropolis graph sampling (MGS), which extensively explore the search space to directly minimize  objectives (see Appendix~\ref{appendix:mgs}).
We used a machine with an i9-10900K CPU and 64GB RAM in all cases except one. 
When running MGS-Avg-Del on the tags-math dataset, we used a machine with an AMD Ryzen 9 3900X CPU and 128GB RAM  due to its large memory requirement.
The sample quality in each method was averaged over three trials.
\subsection{Q1. Quality of Samples}
\label{sec:evaluation:quality}
We compare all 13 considered sampling methods numerically using distances (D-statistics or differences), rankings, and Z-Scores, as described in Sections~\ref{problem_form:problem} and \ref{sec:observations:evaluation}.
The results under five sampling portions (10\%-50\%)
are averaged in Table~\ref{tab:evaluation2}, and some are visualized in Figure~\ref{fig:evaluation:all}. 
\textbf{\methodauto provides overall the most representative samples, among the $13$ considered methods, in terms of both average rankings and average Z-Scores}. Especially, \methodauto best preserves  node degrees, density, overlapness, and  diameter. Compared to \textit{RHS}, the D-statistics in degree distributions drop significantly, and  the differences in density, overlapness, and diameters drop more significantly. 
That is, better preserving node degrees in \methodauto helps resolve the weaknesses of \textit{RHS}.
While \methodauto is outperformed by \textit{RHS} in preserving hyperedge sizes, intersection sizes, relative singular value, and connected component sizes, the gaps between their D-Statistics or differences are less than $0.05$. 



\subsection{Q2. Robustness}
We demonstrate the robustness of \methodauto to sampling portions.
In Figure~\ref{fig:evaluation:all}, we show how D-statistics in degree distributions, average Z-Scores, and average rankings change depending on sampling proportions.
\methodauto is consistently best regardless of sampling portions with few exceptions. \textit{MGS} methods preserve intersection sizes, node-pair degrees, hyperedge sizes, relative single values, and connected component sizes better than \methodauto by small margins, and as a result, \methodauto is outperformed by some MGS methods in terms of average ranking in a few settings.


\subsection{Q3. Speed}
\label{sec:evaluation:speed}
We measured the running times of all considered sampling methods in each dataset with five sampling portions.
We compare the sum of running times in Figure~\ref{fig:evaluation4:time}.
Despite its additional overhead for automatic hyperparameter tuning, \methodauto is the fastest except for the simplest methods (\textit{RHS}, \textit{DNS}, and \textit{RNS}). Notably, it is $2.7\times$ faster than \method with grid search (without any  degradation in sample quality) when the search space for $\alpha$ is fixed.
The speed and sample quality are plotted together in Figure~\ref{fig:strength}.

	\section{Related Work}
	\label{sec:related}
	
\smallsection{Graph Sampling:}
Many studies deal with subgraph sampling with specific objectives, including the preservation of community structures~\cite{maiya2010sampling}, the maximization of classification accuracy \cite{yoo2020sampling}, the attainment of good visualizations~\cite{gilbert2004compressing, rafiei2005effectively, jia2008visualization}, the effective crawling of online social networks~\cite{kurant2011towards,lee2012beyond,xu2014general,li2015random}, and the accurate estimation of triangle counts \cite{lim2015mascot,stefani2017triest,lee2020temporal}.
Sampling a \emph{representative} subgraph that preserves multiple properties of the original graph has also been extensively studied \cite{leskovec2006sampling, hubler2008metropolis, ahmed2011network, voudigari2016rank}. 



In terms of methodologies, most previous graph-sampling methods can be categorized into (a) node selection methods and (b) edge selection methods, as we categorize hypergraph-sampling methods in Section \ref{sec:baseline}.
Node-selection methods \cite{leskovec2006sampling, hubler2008metropolis, maiya2010sampling} first select a subset of nodes and yield the subgraph induced by the subset.
Edge-selection methods \cite{leskovec2006sampling, ahmed2011network,lim2015mascot,stefani2017triest,lee2020temporal} select a subset of edges first and yield the subgraph composed of the edges and all incident nodes.
Node-selection methods are relatively good at preserving node-level properties (e.g., the distribution of node degrees), while edge-selection methods are known to preserve the distribution of sizes of connected components relatively well \cite{leskovec2006sampling, ahmed2011network}.


In this paper, we study the problem of representative hypergraph sampling. We examine several baseline approaches that generalize the aforementioned graph-sampling approaches to the problem, and we demonstrate their previously known properties do not hold in our case. Thus, we develop a novel effective approach that addresses the limitations of the baseline approaches.

\smallsection{Hypergraph Sampling:}
Hypergraphs have drawn a lot of attention lately in various domains, including recommendation \cite{wang2020next}, entity ranking \cite{chitra2019random}, and node classification \cite{feng2019hypergraph, jiang2019dynamic}.
In them, high-order relationships between nodes embedded in hyperedges are exploited to achieve better performance in various tasks.

Despite the growing interest in hypergraphs, the problem of hypergraph sampling has received surprisingly little attention. Yang et al.
\cite{yang2019revisiting} devised a novel sampling method for hypergraphs, but it is specialized for a specific task (specifically, node embedding). 
To the best of our knowledge, there is no previous work that aims at sampling representative sub-hypergraphs that preserve multiple structural properties of the original hypergraph.
There have been studies of generating hypergraphs that mimic the properties of real-world hypergraphs~\cite{deza2019hypergraphic, chodrow2020configuration, arafat2020construction, dyer2020sampling,lee2021hyperedges,do2020structural,kook2020evolution}. In them, new nodes and hyperedges are freely created, while we focus on drawing a subset from existing ones.


With respect to hypergraph sampling, it is important to decide the non-trivial structural properties of hypergraphs to be preserved. 
Kook et al. \cite{kook2020evolution} discovered unique patterns in real-world hypergraphs regarding (a) hyperedge sizes, (b) intersection sizes, (c) the singular values of the incidence matrix, (d) edge density, and (e) diameter. Lee et al. \cite{lee2021hyperedges} reported a number of properties regarding the overlaps of hyperedges by which real-world hypergraphs are clearly distinguished from random hypergraphs.
In addition, Do et al. \cite{do2020structural} uncovered several patterns regarding the connections between subsets of a fixed number of nodes (e.g., connected component sizes) in real-world hypergraphs.


	\section{Conclusions}
	\label{sec:conclusion}
In this work, we propose \methodauto, a fast and effective algorithm for sampling representative sub-hypergraphs.
The key idea was to remedy the limitations of random hyperedge sampling (\textit{RHS}), which has unique merits, by automatically adjusting the amount of bias towards high-degree nodes. 
Future work could consider representative sampling from streaming hypergraphs.
Our contributions are summarized as follows:
\begin{itemize}[leftmargin=*]
    \item \textbf{Problem Formulation}: To the best of our knowledge, we formulate the problem of representative sampling from real-world hypergraphs for the first time. Our formulation is based on ten pervasive structural properties of them. 
    \item \textbf{Observations} : We examine the characteristics of six intuitive sampling approaches in eleven datasets, and our findings guide the development of a better algorithm.
    \item \textbf{Algorithm Design}: We propose \methodauto, and it is fast while sampling overall the best sub-hypergraphs among $13$ methods.
\end{itemize}
\textbf{The code and datasets} are available at \cite{online2021appendix} for reproducibility.


    {\small \smallsection{Acknowledgements} This work was supported by National Research Foundation of Korea (NRF) grant funded by the Korea government (MSIT) (No. NRF-2020R1C1C1008296) and Institute of Information \& Communications Technology Planning \& Evaluation (IITP) grant funded by the Korea government (MSIT) (No. 2019-0-00075, Artificial Intelligence Graduate School Program (KAIST)).}

    \bibliographystyle{ACM-Reference-Format}
	\bibliography{BIB/ref}

    \clearpage
    \newpage
	\appendix
	\section{Appendix}

\begin{table}[t]
    \vspace{-3mm}
	\begin{center}
		\caption{\label{tab:datasets}Summary of real-world hypergraphs.}
		\scalebox{0.66}{
            \begin{tabular}{l| c c c c c c c c c}
				\toprule
				\textbf{Dataset} & \bigcell{c}{$|\SV|$} & \bigcell{c}{$|\SE|$} & \bigcell{c}{AVG.\\d($v$)} & \bigcell{c}{AVG.\\|$e$|} & \bigcell{c}{No. of\\CCs} & \bigcell{c}{Largest\\CC} & \bigcell{c}{GCC} & Density & \bigcell{c}{Diameter} \\ 
				\midrule
				email-Enron & 143 & 1,514 & 32.3 & 3.05 & 1 & 143 & 0.66 & 10.59 & 2.38 \\
				email-Eu & 1,005 & 25,148 & 88.9 & 3.56 & 20 & 986 & 0.57 & 25.02 & 2.78 \\
				\midrule
				contact-primary & 242 & 12,704 & 126.9 & 2.42 & 1 & 242 & 0.53 & 52.50 & 1.88 \\
				contact-high & 327 & 7,818 & 55.6 & 2.33 & 1 & 327 & 0.50 & 23.91 & 2.63 \\
				\midrule
				NDC-classes & 1,161 & 1,090 & 5.6 & 5.97 & 183 & 628 & 0.83 & 0.94 & 4.65\\
				NDC-substances & 5,556 & 10,273 & 12.2 & 6.62 & 1,888 & 3,414 & 0.72 & 1.85 & 3.04 \\
				\midrule
				tags-ubuntu & 3,029 & 147K & 164.8 & 3.39 & 9 & 3,021 & 0.61 & 48.60 & 2.41 \\
				tags-math & 1,629 & 170K & 364.1 & 3.48 & 3 & 1,627 & 0.63 & 104.65 & 2.13 \\
				\midrule
				threads-ubuntu & 125K & 166K & 2.5 & 1.91 & 39K & 82K & 0.55 & 1.33 & 4.73 \\
				\midrule
				coauth-geology & 1.2M & 1.2M & 3.0 & 3.17 & 230K & 903K & 0.76 & 0.96 & 7.04 \\
				coauth-history & 1M & 896K & 1.3 & 1.57 & 617K & 242K & 0.82 & 0.87 & 11.28 \\
				\bottomrule
			\end{tabular}}
	\end{center}
\end{table}

\subsection{Datasets}\label{appendix:datasets}

Throughout the paper, we use $11$ datasets summarized in Table~\ref{tab:datasets} after removing all duplicated hyperedges.
Their domains are:


\noindent $\circ$ \textbf{email} (email-Enron~\cite{klimt2004enron} and email-Eu~\cite{leskovec2005graphs,yin2017local}): Each hyperedge represents an email. It consists of the sender and  receivers.

\noindent $\circ$ \textbf{contact} (contact-primary~\cite{stehle2011high} and contact-high~\cite{mastrandrea2015contact}): Each hyperedge represents a group interaction. It consists of individuals.

\noindent $\circ$ \textbf{drugs} (NDC-classes and NDC-substances): Each hyperedge represents an NDC code for a drug. It consists of  classes or substances.

\noindent $\circ$ \textbf{tags} (tags-ubuntu and tags-math): Each hyperedge represents a post. It consists of tags.

\noindent $\circ$ \textbf{threads} (threads-ubuntu): Each hyperedge represents a question. It consists of answerers.

\noindent $\circ$ \textbf{co-authorship} (coauth-geology~\cite{sinha2015overview} and coauth-history~\cite{sinha2015overview}): Each hyperedge represents a publication. It consists of co-authors.

\subsection{Proof of Theorem~\ref{theorem:time}} \label{appendix:timecomplexity}
\begin{proof}
It takes $O(\sum_{e\in \SE} |e|)$ to compute $\omega(e)$ for every hyperedge $e$, and it takes $O(|\SE|)$ time to build a balanced binary tree with $\max_{e \in \SE}$ $\omega(e)$ leaf nodes where each $k$-th leaf node points to the list of all hyperedges whose weight is $k^{\alpha}$.
Then, 
it takes $O(\max_{e\in \SE} \omega(e))=O(|\SE|)$ time to store in each node $i$ the sum of the weights of the hyperedges pointed by any node in the sub-tree rooted at $i$ if we store them from leaf nodes to the root.
The height of the tree is $O(\log (\max_{e\in \SE} \omega(e)))=O(\log (\max_{v\in \SV} d_{v}))$, and thus
drawing each hyperedge (i.e., from the root, repeatedly choosing a child with weights until reaching a leaf; and then drawing a hyperedge that the leaf points to) and updating weights accordingly takes $O(\log (\max_{v\in \SV} d_{v}))$ time.
Drawing $p \cdot |\SE|$ hyperedges takes $p \cdot |\SE| \cdot \log(\max_{v\in \SV} d_{v}))$ time, and since $O(|\SE|)=O(\sum_{e\in \SE} |e|)$, the total time complexity is $O(p \cdot |\SE| \cdot \log(\max_{v\in \SV} d_{v})+\sum_{e\in \SE} |e|)$.
\end{proof}


\subsection{Metropolis Graph Sampling (MGS)}\label{appendix:mgs}

Below, we describe the \textit{MGS} methods, which are the strongest competitors in Section~\ref{sec:evaluation:quality}.
They adopt metropolis graph sampling~\cite{hubler2008metropolis} for Problem~\ref{problem}. For a sub-hypergraph $\SGH$, $\varrho^{*}(\SGH)$ := $\frac{1}{\exp(k \cdot \Delta_{\SG}(\SGH))}$. Then, the acceptance probability of a move from a state $\SGH$ to a state 
 ${\SGH}'$ is 
min(1,$\frac{\varrho^{*}({\SGH}')}{\varrho^{*}(\SGH)}$) = min(1, $\exp(k \cdot (\Delta_{\SG}(\SGH) - \Delta_{\SG}({\SGH}'))$). This algorithm makes greedy choices that decrease a predefined objective function $\Delta_{\SG}(\SGH)$ most. In our experiments, we used the $k$ values in the the search space $\{1, 10, 100, 10000\}$.

Depending on $\Delta_{\SG}(\SGH)$, we divide \textit{MGS} into \textit{MGS-Deg} and \textit{MGS-Avg}.
The former aims to preserve node degrees by setting $\Delta_{\SG}(\SGH)$ to the D-statistic between degree distributions in $\SG$ and $\SGH$. The latter aims to preserve node degrees, hyperedge sizes, node-pair degrees, and hyperedge intersection sizes at the same time by setting $\Delta_{\SG}(\SGH)$ to the average of the D-statistics from their distributions.
The four statistics are chosen since they are cheap to compute at every step.

We further divide each of \textit{MGS-Deg} and \textit{MGS-Avg} to three versions depending on how to move between states as follows:




\begin{itemize}[leftmargin=*]
\item \textbf{Add}: MGS-Deg-Add (\textit{MGS-DA}) and MGS-Avg-Add (\textit{MGS-AA}) start from $\SEH=\emptyset$ and repeatedly propose to add a hyperedge to $\SEH$ until $|\SEH| = \lfloor |\SE| \cdot p \rfloor$ holds.
    
\item \textbf{Replace}: MGS-Deg-Rep (\textit{MGS-DT}) and MGS-Avg-Rep (\textit{MGS-AR}) initialize $\SEH$ so that $|\SEH| = \lfloor |\SE| \cdot p \rfloor$ by using \textit{RHS}. They repeat $3000$ times proposing to replace a hyperedge in $\SEH$ with one outside $\SEH$.


\item \textbf{Delete}: MGS-Deg-Del (\textit{MGS-DD}) and MGS-Avg-Del (\textit{MGS-AD}) start from $\SEH = \SE$ and repeatedly propose to remove a hyperedge from $\SEH$ until $|\SEH| = \lfloor |\SE| \cdot p \rfloor$ holds.
\end{itemize}



\begin{table*}[t]
\vspace{-3mm}
\caption{\label{tab:ablation_eval}
\method gives overall more representative samples than its three variants. 
We report rankings and Z-Scores (in parentheses) averaged over all $\bf{11}$ datasets and $\bf{5}$ different sampling portions (10\%, 20\%, 30\%, 40\%, and 50\%).}
    \begin{center}
    \scalebox{0.74}{
        \begin{tabular}{c cccccccccc | c}
            \toprule
            \textbf{Algorithm} & \textbf{Degree} & \textbf{Int. Sizes.} & \textbf{Pair Degree} & \textbf{Size} & \textbf{SV} & \textbf{CC} & \textbf{GCC} & \textbf{Density} & \textbf{Overlapness} & \textbf{Diameter} & \textbf{AVG} \\
            
            
            
            \hline
            {\method} & \textbf{1.91} (\textbf{-0.45}) & 2.27 (-0.06) & 2.56 (0.09) & \textbf{1.69} (\textbf{-0.73}) & 2.16 (-0.07) & \textbf{1.65} (\textbf{-0.32}) & \textbf{1.87} (\textbf{-0.41}) & \textbf{1.62} (\textbf{-0.62}) & \textbf{1.91} (\textbf{-0.43}) & \textbf{1.82} (\textbf{-0.46}) & \textbf{1.95} (\textbf{-0.35}) \\
            {\method-\textsc{Avg}} & 2.62  (0.24) & 2.07  (-0.18) & \textbf{2.04} (\textbf{-0.32}) & 2.33 (0.06) & \textbf{1.75} (\textbf{-0.24}) & 2.04 (0.12) & 2.24 (0.05) & 2.69 (0.39) & 2.55 (0.28) & 2.27 (-0.03) & 2.26 (0.04) \\
            {\method-\textsc{NS}} & 2.15 (-0.20) & 3.27 (0.57) & 2.85 (0.36) & 2.55 (0.14) & 2.82 (0.41) & 1.80 (-0.19) & 2.87 (0.34) & 2.00 (-0.42) & 2.09 (-0.31) & 2.80 (0.24) & 2.52 (0.09) \\
            {\method-\textsc{Max}} & 2.85 (0.40) & \textbf{1.91} (\textbf{-0.34}) & 2.07 (-0.13) & 2.96 (0.53) & 2.04 (-0.05) & 2.47 (0.39) & 2.55 (0.01) & 3.20 (0.65) & 2.98 (0.46) & 2.64 (0.25) & 2.57 (0.22) \\
            \bottomrule
        \end{tabular}}
    \end{center}
    \vspace{-2mm}
\end{table*}

\begin{figure}[!t]
    \vspace{-3mm}
    \centering
    \begin{minipage}{0.87\linewidth}
        \includegraphics[width=1.0\textwidth]{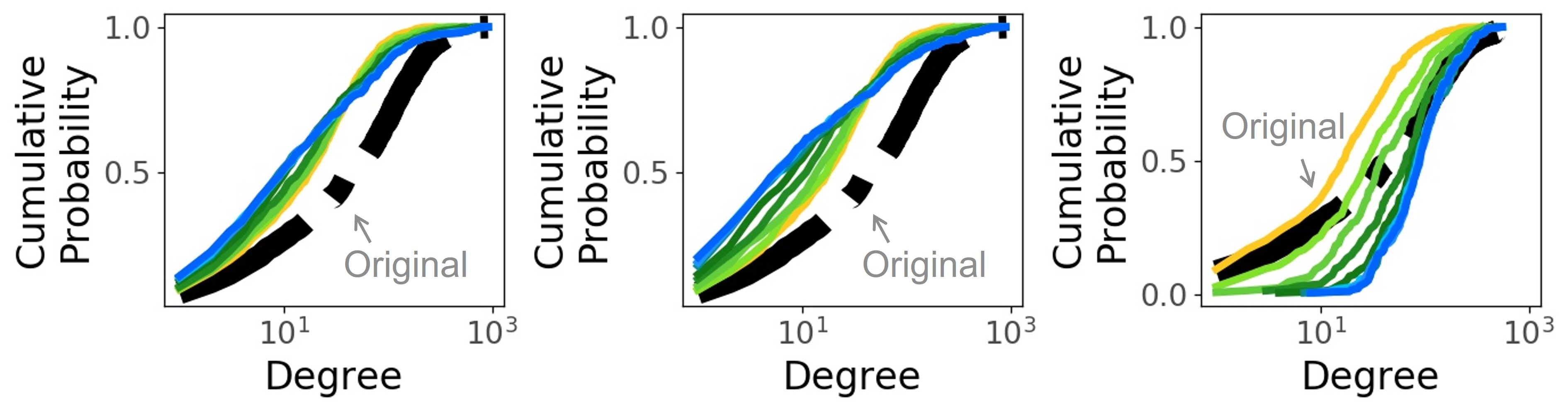} 
    \end{minipage}
    \begin{minipage}{0.09\linewidth}
        \includegraphics[width=0.8\textwidth]{FIG/observation/observation1/alpha_colorbar.jpg}
        \vspace{1mm}
    \end{minipage}
    
    \vspace{-0.5mm}
    
    \begin{minipage}{1.0\linewidth} \small
        \hspace{0.01\linewidth}(a) \method-\textsc{Max}
        \hspace{0.01\linewidth}(b) \method-\textsc{Avg}
        \hspace{0.01\linewidth}(c) \method-\textsc{NS}
    \end{minipage}
    
    \caption{In \method-\textsc{Max} and \method-\textsc{Avg}, 
    the biases of degree distributions in samples cannot be fully controlled by $\alpha$.
    We report the results when sampling 30\% of hyperedges from the email-Eu dataset.}
    \label{fig:appendix:ablation:obs1}
\end{figure}

\begin{figure}[!t]
    \vspace{-3mm}
    \centering
    \begin{minipage}[c]{0.95\linewidth} 
        \includegraphics[width=1.0\textwidth,]{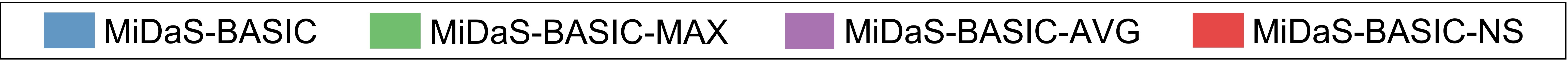}
    \end{minipage}

    \begin{minipage}{0.28\linewidth}
        \includegraphics[width=1\textwidth]{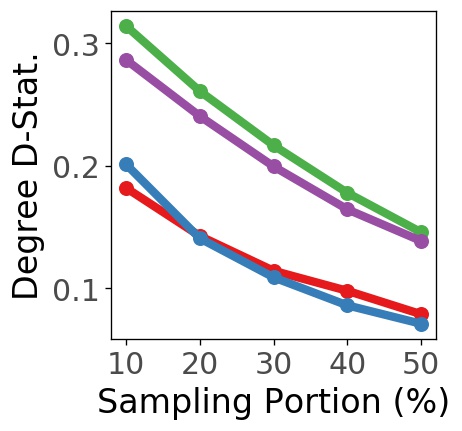}
    \end{minipage}
    \begin{minipage}{0.28\linewidth}
        \includegraphics[width=1\textwidth]{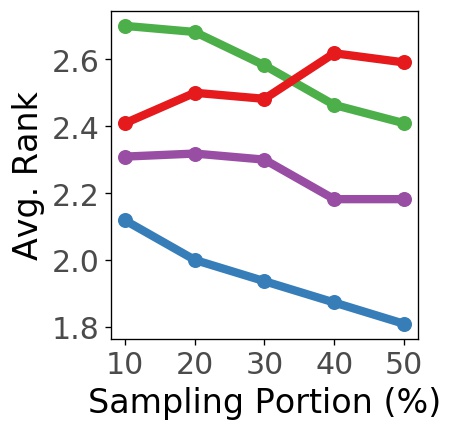}
    \end{minipage}
    \begin{minipage}{0.3\linewidth}
        \includegraphics[width=1\textwidth]{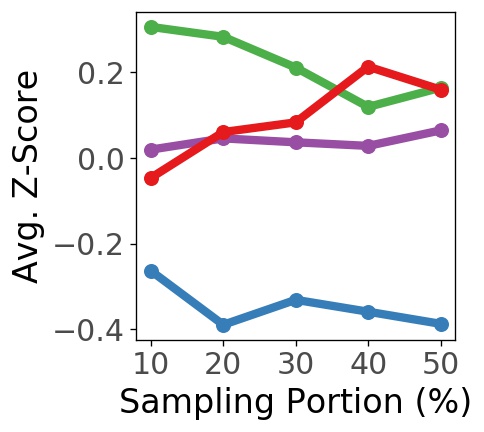}
    \end{minipage}

    \vspace{-0.5mm}
    
    \begin{minipage}{1.0\linewidth} \small
        \hspace{0.05\linewidth}(a) Degree Preservation
        \hspace{0.02\linewidth}(b) Avg. Z-Score
        \hspace{0.08\linewidth}(c) Avg. Ranking
    \end{minipage}
    
    \caption{\method consistently outperforms its variants, in terms of degree preservation, average rankings, and average Z-Scores. The results justify our design choices.}
    \label{fig:appendix:ablation:result}
\end{figure}


\subsection{Ablation Study}~\label{appendx:ablation}
Below, in order to justify the design choices that we make when designing \method, we compare it with its three variants:


\begin{itemize}[leftmargin=*]
\item \textbf{\method-\textsc{Max}}: This variant uses $(\max_{v \in e}d_{v})^\alpha$ for hyperedge weighting.
    
\item \textbf{\method-\textsc{Avg}}: This variant uses  $(\text{avg}_{v \in e} d_{v})^\alpha$  for hyperedge weighting.
    
\item \textbf{\method-\textsc{NS}}: This variant draws nodes with probability proportional to $\{d_{v}\}^{\alpha}$ and returns the induced sub-hypergraph (see Section \ref{sec:baseline:NS}).
\end{itemize}

We first examine whether the variants also exhibit  Observation~\ref{obs:alpha_and_dist}. As shown in Figure~\ref{fig:appendix:ablation:obs1}, only \method-\textsc{NS} exhibits it. In \method-\textsc{Max} and \method-\textsc{Avg}, 
the biases of degree distributions in samples change within a limited range.

Then, we compare \method with the three variants in terms of (a) D-statistics between degree distributions, (b) average rankings (over all the datasets), and (c) average Z-Scores (over all the datasets), after tuning their $\alpha$ values by grid search in $\{0, 2^{-1}, 2^{0}, ..., 2^{6}\}$.
As shown in Figure~\ref{fig:appendix:ablation:result}(a), both \method-\textsc{Max} and \method-\textsc{Avg} produce degree distributions quite different from the original one; 
while \method-\textsc{NS} preserves the degree distribution accurately.
Overall, all three variants are outperformed by \method, as seen in Table~\ref{tab:ablation_eval} and Figure~\ref{fig:appendix:ablation:result}(b-c).

\section{Theoretical Analysis}\label{appendix:theoretical_analysis}
Below, we present a theoretical analysis of the relation between hyperedge weighting and biases of degree distributions in samples.



\begin{defn}
	Let $S$ be a hyperedge sampling algorithm and $\phi_S(e) \geq 0$ be the weight of a hyperedge $e$ for being selected by $S$.
	Then, we define the probability $p_S(e)$ of $e$ being selected by $S$ as
	\begin{equation*}
		p_S(e) = \frac{\phi_S(e)^\alpha}{\sum_{e' \in \SE}\phi_S(e')^\alpha} = \frac{1}{Z_S(\alpha)} \phi_S(e)^\alpha,
	\end{equation*}
	where $Z_S(\alpha)$ is the normalization constant, and $\alpha$ is a parameter.
\end{defn}

\begin{defn}~\label{def:l_S}
	Given a sampling algorithm $S$, we denote by $l_S(k)$ the probability of sampling a hyperedge that contains a node whose degree is lower than or equal to $k$ from $\SG=(\SV,\SE)$. That is,
	\begin{align*}
		l_S(k) = \sum\nolimits_{e \in \SE(\SV_{k})} p_S(e)
	\end{align*}
	where $\SV_{k} = \{ v \in \SV : d(v) \leq k \}$ and $\SE({A}) = \{e \in \SE : e \cap A \neq \emptyset \}$.
\end{defn}

Using $l_S(k)$, we can define the probability of sampling a hyperedge that contains \textit{only} nodes whose degrees are higher than $k$. We denote this as $h_S(k) = 1 - l_S(k)$. In this definition, $h_S(k)$ indicates the following meaning:

\begin{defn} \label{def:higher_h_k}
	For any $k\geq 0$, if  $h_{A}(k) < h_{B}(k)$, we say Algorithm $B$ is more biased towards nodes with degree higher than $k$, compared to Algorithm $A$.
\end{defn}
Selecting a node with a degree higher than $k$ can be divided into two cases: (a) selecting a hyperedge where at least one node has a degree higher than $k$ but not all (i.e., $\{e \in \SE : \exists v \in e \text{ such that } d(v)  \leq k \text{ and } \exists v' \in e \text{ such that } d(v')  > k \}$) and (b) selecting a hyperedge where all nodes have degrees higher than $k$ (i.e., $\{e \in \SE : \forall v \in e, d(v)  > k \}$). Because the former case increases the probability of sampling a node with a degree less than or equal to $k$, the latter contributes more to being strongly biased towards nodes with degree more than $k$.


Below, we are going to use $M_{\omega}(\alpha)$ to refer to random hyperedge sampling with $\omega(e)^{\alpha}$ as the hyperedge weight function.  
Note that the case of $\alpha = 0$ (i.e., $M_{\omega}(0)$) corresponds to \textit{RHS}; and the case of $\omega(e)=\min_{v \in e} d_{v}$ corresponds to \method.

\begin{thm} \label{theorem:increasing_h}
    Given $k$, $h_{M_{\omega}(\alpha)}(k)$ is an increasing function of $\alpha$, if $k$ satisfies 
    \begin{equation} \label{theorem:eq:em_condition}
        \mathbb{MAX}_{e \in \SE(\SV_{k})} \ln \omega(e) < \mathbb{AVG}_{e \in \SE} \ln \omega(e).
    \end{equation}
\end{thm}

\begin{proof}
    $h_{M_{\omega}(\alpha)}(k)$ is an increasing function of $\alpha$ if $\frac{\partial l_{M_{\omega}(\alpha)}(k)}{\partial \alpha}<0$ holds for any $\alpha$.
    $\frac{\partial l_{M_{\omega}(\alpha)}(k)}{\partial \alpha}$ is arranged as
    \begin{equation} \label{theorem:eq:condition}
        \sum_{e \in \SE(\SV_{k})}
            \frac{\omega(e)^{\alpha} \ln \omega(e)}{ \sum_{e' \in \SE(\SV_{k})} \omega(e')^{\alpha}} <
        \sum_{e \in \SE}
            \frac{\omega(e)^{\alpha} \ln \omega(e)}{\sum_{e' \in \SE} \omega(e')^{\alpha}}.
    \end{equation}
    
    For any set of hyperedges $\SEH$, the following equation is satisfied if $\alpha$ changes from 0 to $\infty$:
    \begin{equation}\label{theorem:eq:range}
        \mathbb{AVG}_{e \in \SEH} \ln \omega(e)
        \leq \sum_{e \in \SEH} \frac{\omega(e)^{\alpha} \ln \omega(e)}{ \sum_{e' \in \SEH} \omega(e')^{\alpha}} 
        \leq \mathbb{MAX}_{e \in \SEH} \ln \omega(e).
    \end{equation}
    
    Based on Eq.~\eqref{theorem:eq:range}, we can get the upper bound of the left hand side and the lower bound of the right hand side of Eq.~\eqref{theorem:eq:condition}.

    Thus, if Eq.~\eqref{theorem:eq:em_condition} is satisfied for given $k$, $\frac{\partial l_{M_{\omega}(\alpha)}(k)}{\partial \alpha}<0$ holds for any $\alpha$.
    That is, $h_{M_{\omega}(\alpha)}(k)$ is an increasing function of $\alpha$.
\end{proof}

We have the following corollary and lemma from Theorem~\ref{theorem:increasing_h}.
\begin{cor}
    For all $k$ that satisfies Eq.~\eqref{theorem:eq:em_condition}, \method is more biased toward nodes with degrees larger than $k$ as $\alpha$ increases.
\end{cor}

\begin{lma} \label{lemma:using_max}
    Given any sampling algorithm, if $k=k'$ satisfies the condition of Eq.~\eqref{theorem:eq:em_condition}, then $k < k'$ also satisfies the condition.
\end{lma}

\begin{proof}
    The proof is straightforward as the left hand side of Eq.~\eqref{theorem:eq:em_condition} is an increasing function of $k$.
\end{proof}

\begin{figure}[h]
    \vspace{-4mm}

    \begin{minipage}{1.0\linewidth} \small
        \hspace{0.12\linewidth} \includegraphics[width=0.8\textwidth]{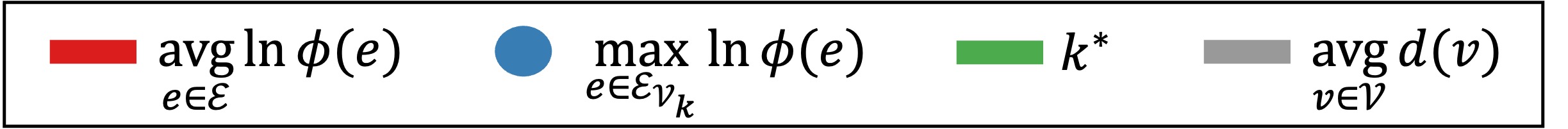}
        \vspace{0.5mm}
    \end{minipage}
    
    \begin{minipage}{1.0\linewidth} \small
        \hspace{0.03\linewidth} \textbf{\method-\textsc{Max}}
        \hspace{0.08\linewidth} \textbf{\method}
        \hspace{0.08\linewidth} \textbf{\method-\textsc{Avg}}
    \end{minipage}
    \begin{minipage}[c]{1.0\linewidth}
        \includegraphics[width=1\textwidth]{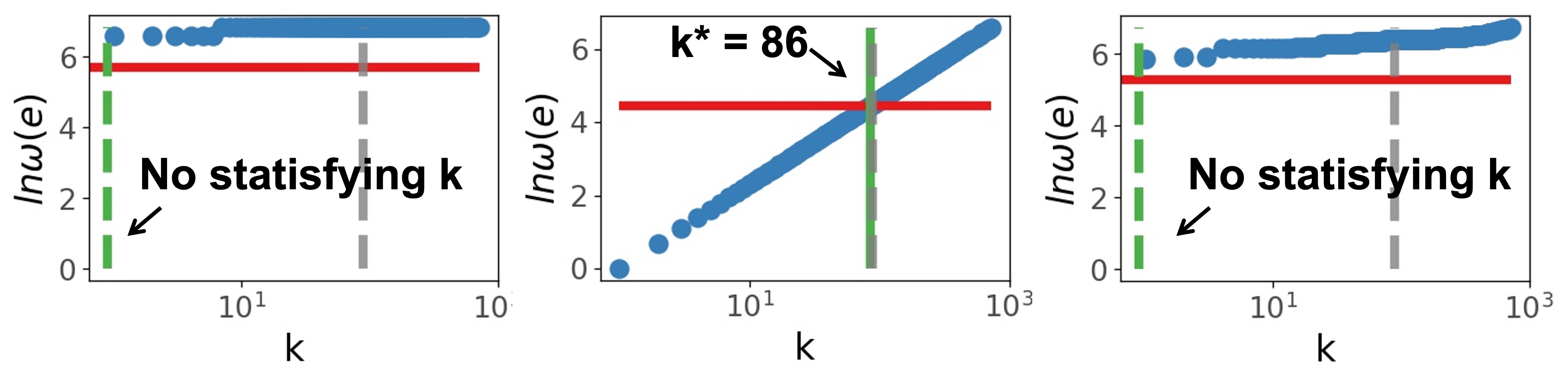}
    \end{minipage}
    
    
    
    
    \vspace{-1mm}
    
    \caption{ \label{fig:em_condition} Eq.~\eqref{theorem:eq:em_condition} is satisfied more easily (i.e., it is satisfied for a wider range of $k$ values) in \method than in its variants. 
    Note that Eq.~\eqref{theorem:eq:em_condition} is a sufficient condition for bias towards high-degree nodes to grow as $\alpha$ increases.}

    \vspace{-3mm}
\end{figure} 

We analyze \method (i.e., $\omega(e)=\min_{v \in e} d_{v}$), \method-\textsc{Max} (i.e., $\omega(e)=\max_{v \in e} d_{v}$), and \method-\textsc{Avg} (i.e., $\omega(e)=\mathrm{avg}_{v \in e} d_{v}$), which are described in Appendix \ref{appendx:ablation}. 
Specifically, we examine whether they have $k$ satisfying Eq.~\eqref{theorem:eq:em_condition} in the $11$ real-world hypergraphs. Based on Lemma~\ref{lemma:using_max}, we examine $k^{*}$, i.e., the maximum $k$ value satisfying Eq.~\eqref{theorem:eq:em_condition}. We find out that \method has satisfying $k$ in all datasets, 
and the result in the Email-Eu dataset is shown in Figure~\ref{fig:em_condition}.
However, both \method-\textsc{Max} and \method-\textsc{Avg} do not have any $k$ satisfying Eq.~\eqref{theorem:eq:em_condition} in most datasets. Even if they have appropriate $k$ in some datasets,  $k^{*}$ values from them are less than those from \method, as summarized below.


\begin{obs}
    On all eleven considered real-world hypergraphs, Eq.~\eqref{theorem:eq:em_condition} is satisfied for a wider range of $k$ values in \method than in its variants.
    Recall that Eq.~\eqref{theorem:eq:em_condition} is a sufficient condition for bias towards high-degree nodes grows as $\alpha$ increases.
\end{obs}

\end{document}